\documentclass[conference]{IEEEtran}
\usepackage{amsfonts}
\IEEEoverridecommandlockouts

\ifCLASSINFOpdf
\else
\fi
\usepackage{epsfig}
\usepackage{graphicx}
\usepackage{psfig}
\usepackage{epsf}
\usepackage[cmex10]{amsmath}
\usepackage{booktabs}
\usepackage{fancyhdr}
\usepackage{slashbox}
\usepackage{multirow}
\usepackage{caption2}   
\usepackage{color}
\newtheorem{myTheo}{Theorem}   

\newtheorem{lemma}{\text{Lemma}}
\newtheorem{rem}{Remark} 

\begin{document}
\title{Computation Efficiency Maximization in Wireless-Powered Mobile Edge Computing Networks}
\author{\IEEEauthorblockN{Fuhui Zhou, \emph{Member, IEEE}, and Rose Qingyang Hu, \emph{Fellow, IEEE}}
\thanks{This paper was presented in part at the IEEE Global Communications Conference (GLOBECOM), Abu Dhabi, UAE, December 2018 \cite{F. H. Zhou}.

Manuscript received March 3, 2019; revised Aug. 6, 2019 and Dec. 16, 2019; accepted January 21, 2019. Date of publication February ******; date of current version ******. This work was supported by the Natural Science Foundation of China under Grants 61701214, in part by The Open Foundation of The State Key Laboratory of Integrated Services Networks under Grant ISN19-08, in part by the Intel Corporation, and in part by the US National Science Foundation under Grant EARS1547312, in part by the Excellent Youth Foundation of Jiangxi Province under Grant 2018ACB21012 and in part by Young Elite Scientist Sponsorship Program by CAST. The associate editor coordinating the review of this paper and approving it for publication was Prof. Jemin Lee. (\emph{Corresponding authors: Fuhui Zhou})

Fuhui Zhou is with the College of Electronic and Information Engineering, Nanjing University of Aeronautics and Astronautics, Nanjing, 210000, P. R. China, and also with the State Key Laboratory of Integrated Services Networks, Xidian University, Xian 710071, China (e-mail: zhoufuhui@ieee.org).

Rose Qingyang Hu is with the Department of Electrical and Computer Engineering, Utah State University, USA.  (e-mail: rose.hu@usu.edu).
}

}
\maketitle
\begin{abstract}
Energy-efficient computation is an inevitable trend for mobile edge computing (MEC) networks. Resource allocation strategies for maximizing the computation efficiency are critically important. In this paper, computation efficiency maximization problems are formulated in wireless-powered MEC networks under both partial and binary computation offloading modes. A practical non-linear energy harvesting model is considered. Both time division multiple access (TDMA) and non-orthogonal multiple access (NOMA) are considered and evaluated  for offloading. The energy harvesting time, the local computing frequency, and the offloading time and power are jointly optimized to maximize the computation efficiency under the max-min fairness criterion.  Two iterative algorithms and two alternative optimization algorithms are respectively proposed to address the non-convex problems formulated in this paper.  Simulation results show that the proposed resource allocation schemes outperform the benchmark schemes in terms of user fairness.  Moreover, a tradeoff is elucidated between the achievable computation efficiency and the total number of computed bits. Furthermore, simulation results demonstrate that the partial computation offloading mode outperforms the binary computation offloading mode and NOMA outperforms TDMA in terms of computation efficiency.
\end{abstract}
\begin{IEEEkeywords}
Mobile-edge computing,  wireless power transfer, computation efficiency, resource allocation, binary computation offloading, partial computation offloading.
\end{IEEEkeywords}
\IEEEpeerreviewmaketitle
\section{Introduction}
\subsection{Background and Related Works}
The emerging intelligent applications (e.g., automatic navigation, face recognition, unmanned driving, etc.) have imposed great challenges for mobile devices since most of those applications have computation-intensive and latency-sensitive tasks to be executed \cite{F. H. Zhou}, \cite{Y. Mao}. However, most mobile devices have low computing capability and finite battery capacity.  Mobile edge computing (MEC)  can significantly augment the computing capability of mobile devices by offloading tasks from mobile devices to the nearby MEC servers in a low latency manner \cite{Y. Mao}.  There exist two operation modes in MEC networks, namely, partial computation offloading and binary computation offloading. In the partial offloading mode, computation tasks can be divided into two parts, one part is executed locally at mobile devices and the other part is offloaded to the MEC server for computing. For the binary computation offloading mode, computation tasks cannot be partitioned.  The entire task is either locally executed or completely offloaded to the MEC server for computing \cite{Y. Mao}, \cite{Y. Wu11}. In this paper, both modes are considered.

To further improve the energy efficiency, wireless powered techniques that exploit radio frequency (RF) signals as the energy sources for powering the energy-limited mobile devices are considered promising and viable approaches in MEC networks since they can provide stable and controllable amount of energy and prolong the battery life of mobile devices \cite{E. Boshkovska2}, \cite{F. Wang}. Recently, an increasing attention has been paid to the wireless powered MEC networks  \cite{F. Wang}. It was shown that user quality of experience (QoE) can be improved by integrating wireless powered techniques into MEC networks since the duration of having MEC services is extended  \cite{S. Bi}. However, due to the ever-increasing greenhouse gas emission concerns and the rapid growth of the operational cost, future MEC networks will more and more focus on maximizing the computation efficiency (CE) \cite{F. Zhou2}, which is defined as the ratio of the total computed bits to the consumed energy. According to \cite{F. Zhou2}, information and communication technologies account for about 2$\%$ of the greenhouse gas and 2$\%$ to 10$\%$ of global energy consumption. In order to achieve a sustainable and green operation of MEC networks, it is crucial to design resource allocation  strategies for maximizing CE of MEC networks. To the authors' best knowledge, there have been only  a few studies in this area. The related works are summarized as follows.

The related works can be classified into three categories.  The first category has focused on designing energy-efficient resource allocation schemes in the conventional MEC networks with orthogonal multiple access (OMA) in \cite{Y. Wang}-\cite{J. Xu1} or with NOMA in \cite{Z. Ding}-\cite{J. Xu3}.  In the second category   resource allocation strategies have been designed  in wireless powered MEC networks \cite{F. Wang}, \cite{S. Bi}, \cite{C. You}-\cite{F. Zhou1}. The third category  has designed energy-efficient resource allocation schemes in wireless-powered networks relying on either OMA or NOMA \cite{Q. Wu}-\cite{T. A. Zewde}.
\subsubsection{Energy-Efficient Resource Allocation in The Conventional MEC Networks}
In the OMA-based MEC networks, efforts have been dedicated to jointly optimizing the communication and computation resources for achieving energy-efficient computing. Specifically, in \cite{Y. Wang}, the consumed energy and execution latency were minimized by jointly optimizing the local computing frequency and offloading power of users.  The authors in \cite{X. Tao} and \cite{C. You1} extended the energy consumption minimization problems into multi-user MEC networks with TDMA and orthogonal frequency-division multiple access (OFDMA), respectively. In \cite{X. Tao}, the computation performance of each user was guaranteed by optimizing the offloading computation bits ratio and offloading time. In \cite{C. You1},  it was shown that the users with strong channel state information (CSI) prefer to offload their computation task to the MEC server while  users with weak CSI chooses to perform local computing. In order to further reduce the energy consumption, the authors in \cite{W. Zhang} studied the coexistence of MEC and cloud computing servers and proposed optimal scheduling policies. Different form the works in \cite{Y. Wang}-\cite{W. Zhang},  multi-antenna techniques were exploited to improve the offloading efficiency \cite{S. Sardellitti}, \cite{T. T. Nguyen}. The energy was minimized by jointly optimizing the beamforming vector and the computation frequency. Recently, the authors in \cite{J. Xu1}  studied secure offloading in multi-user MEC networks. The works in \cite{Y. Wang}-\cite{J. Xu1}  focused on optimizing a single objective, which cannot achieve a good tradeoff among different performance metrics. In \cite{F. Zhou2}, the authors studied CE maximization problem in an MEC system with TDMA, which achieves a good tradeoff between the achievable computation bits and the energy consumption.

Recently, in order to increase connectivity and reduce access latency, resource allocation problems were studied in MEC networks with NOMA \cite{Z. Ding}-\cite{J. Xu3}. In \cite{Z. Ding}, the impact of NOMA on offloading was analyzed in MEC networks. It was proved that the application of NOMA can efficiently reduce the energy consumption and offloading delay compared to OMA. The authors in \cite{A. Kiani}-\cite{J. Xu3}  designed resource allocation strategies for minimizing the consumption energy of various MEC networks with NOMA. Specifically,  the user clustering, communication and computing resources were jointly optimized in multi-cell MEC networks in \cite{A. Kiani} while the communication and computing resource and the trajectory of the unmanned aerial vehicle (UAV) were jointly optimized in \cite{S. Jeong}. In \cite{J. Xu3}, the authors  studied a multi-antenna MEC network with NOMA. It was shown that the exploitation of multi-antenna techniques can significantly improve the offloading efficiency.
\subsubsection{Resource Allocation in Wireless Powered MEC Networks}
Recently, the authors in \cite{F. Wang}, \cite{S. Bi}, \cite{C. You}-\cite{F. Zhou1} studied resource allocation problems in various wireless powered MEC networks. Specifically, in  \cite{C. You}, the central processing unit (CPU) frequency of the user and the mode selection were  jointly optimized for minimizing the consumed energy under both  causal and non-causal CSI conditions. The reinforcement learning and Lyapunov optimization theory were exploited to design resource allocation strategies for minimizing the system cost of wireless powered MEC networks in \cite{J. Xu} and \cite{Y. Mao1}, respectively. Although the computation performance can be improved by integrating wireless powered techniques into MEC networks \cite{C. You}-\cite{Y. Mao1}, the performance improvement is limited by the harvested energy. In order to improve the energy conservation efficiency, the authors in  \cite{F. Wang}  exploited multi-antenna techniques and jointly optimized the energy transmit beam-former, the CPU frequencies and the offloading bits for minimizing the total energy consumption. In \cite{X. Hu}, the authors extended the energy consumption minimization problem into a cooperation-assisted wireless powered MEC network.  Different from the works in \cite{F. Wang}, \cite{C. You}-\cite{Y. Mao1}, the authors in \cite{S. Mao} and \cite{S. Mao1}  proposed optimal resource allocation strategies for maximizing the CE of multi-user and full-duplex wireless powered MEC networks. In contrast to the works in \cite{F. Wang}, \cite{C. You}-\cite{S. Mao1}, resource allocation strategies were designed for maximizing the computation bits of multi-user and UAV-enabled wireless powered MEC systems in \cite{S. Bi} and \cite{F. Zhou1}  under the binary computation offloading mode.
\subsubsection{Energy-Efficient Resource Allocation in Wireless Powered Networks} In conventional wireless powered networks where the computing process was not considered, the energy efficiency maximization problems have been well studied in \cite{Q. Wu}-\cite{T. A. Zewde}. Specifically, the authors in \cite{Q. Wu} and \cite{Q. Wu2} designed resource allocation strategies for maximizing the system  and user-centric energy efficiency, respectively. In order to improve the achievable energy efficiency, multiple-input multiple-out (MIMO) and massive MIMO techniques have been applied in wireless powered networks in \cite{T. A. Khan} and \cite{Z. Chang}, respectively. Different from the works in \cite{Q. Wu}-\cite{Z. Chang} where TDMA was applied for serving multiple users, the authors in \cite{Q. Wu3} and \cite{T. A. Zewde} have designed energy-efficient resource allocation schemes in the wireless powered networks with NOMA. It was interesting to find that NOMA does not necessarily guarantee to achieve a better energy efficiency compared to TDMA.
\subsection{Motivations and Contributions}
Note that the resource allocation strategies proposed in the conventional MEC networks \cite{Y. Wang}-\cite{J. Xu3} cannot be readily applied to wireless powered MEC systems, where the resource allocation problems should consider the EH causal constraints and the relationship among the EH, offloading and computing process. Moreover,  the resource allocation strategies proposed for wireless powered  MEC networks in \cite{F. Wang}, \cite{S. Bi}, \cite{C. You}-\cite{F. Zhou1} are based on an ideally linear EH model. These schemes may not achieve good performance in reality as the practical EH circuits can result in a non-linear  end-to-end wireless power conversion \cite{S. Wang}-\cite{X. Zhang}. Furthermore, resource allocation strategies designed for maximizing the computation bits under the binary computation offloading mode \cite{S. Bi} and \cite{F. Zhou1}  and designed for maximizing CE  under partial offloading mode cannot guarantee to maximize the CE under the binary computation offloading mode. Additionally, although energy-efficiency resource allocation strategies have been designed in the conventional wireless powered networks \cite{Q. Wu}-\cite{T. A. Zewde}, only the efficiency of the computation task offloading process, i.e., communications  process, was considered. They are inappropriate in wireless-powered MEC networks for maximizing the CE since energy consumed in both the offloading and local computation processes should be considered.

To the authors' best knowledge, this is the first work that comprehensively studies resource allocation problems for maximizing CE under both  partial and binary computation offloading modes. Please note that in \cite{F. H. Zhou}, we only studied the CE  problem   under the partial offloading mode with TDMA. Moreover, we did not consider the effect of the power amplifier coefficient on CE  \cite{F. H. Zhou}. The main contributions of our work in this paper are summarized as follows.

\begin{enumerate}
  \item It is the first time that the CE maximization framework is formulated in wireless powered MEC networks under both partial and binary computation offloading modes. Both TDMA and NOMA are considered for offloading transmission. With TDMA, the closed-form expressions for the optimal CPU frequency and the optimal offloading power of users are derived under the partial offloading mode and the close-form expression for the optimal operational mode selection is given under the binary computation offloading mode. An iterative algorithm and an alternative optimization algorithm are proposed to solve the CE  maximization under the partial offloading and binary computation offloading mode, respectively.
  \item With NOMA,  an iterative algorithm and an alternative optimization algorithm based on the  successive convex approximation (SCA) method are proposed for the partial offloading mode and the binary computation offloading mode, respectively. The closed-form expression for the operational mode selection on whether users choose to locally compute or to offload tasks is derived. It is shown that the selection of the operational mode depends on the trade-off between the achievable computation bits and the energy consumption cost of users.
  \item Simulation results show that our proposed resource allocation strategies can improve  fairness among users in terms of CE compared to the benchmark scheme. Moreover, it is shown that the partial offloading mode and NOMA can achieve CE gains compared with the binary computation offloading mode and TDMA, respectively. Furthermore, the tradeoff between the achievable CE  and the computation bits is firstly elucidated.
\end{enumerate}

The remainder of this paper is organized as follows. Section II presents the system model. $\text{C}_{\text{eff}}$  maximization problems are investigated for wireless powered MEC networks with TDMA in Section III. Section IV presents the CE maximization problems in wireless powered MEC networks with NOMA. Simulation results are presented in Section V. The paper is concluded in Section VI.
\section{System Model}
\begin{figure}[!t]
\centering
\includegraphics[width=3.0in]{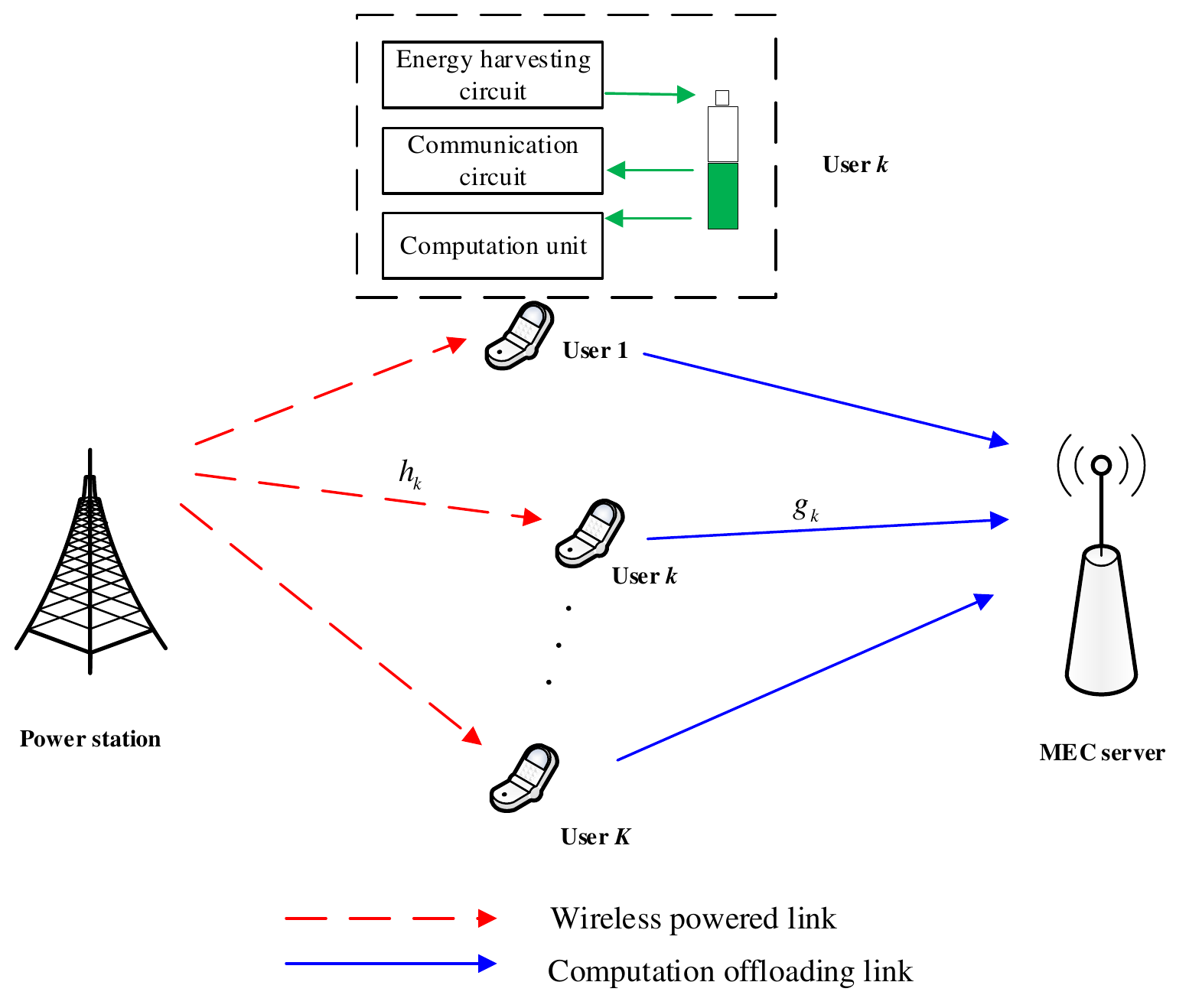}
\caption{The system model.} \label{fig.1}
\end{figure}
A wireless powered MEC network is considered in Fig. 1, where the wireless power station provides the wireless power transfer (WPT) services for $K$ users. Similar to the works in \cite{S. Bi}, \cite{C. You} and \cite{J. Xu}, in order to clarify the issues pertaining to CE and permit reaching meaningful insights into the CE maximization problem, it is assumed that all devices are equipped with a single antenna. In this paper, both partial and binary computation offloading modes are considered. Similar to \cite{F. Wang}, \cite{S. Bi}, \cite{Y. Mao1}-\cite{F. Zhou1}, local computation and downlink WPT can be simultaneously executed while the downlink WPT and the uplink computation offloading cannot be simultaneously performed. Thus, a harvest-then-offload protocol is applied for downlink WPT and uplink computation offloading. Moreover,  both TDMA and NOMA protocols are exploited for achieving multi-user offloading during the offloading process. All the nodes and devices are equipped with a single antenna. Similar to \cite{C. You}-\cite{F. Zhou1}, all the channels have block fading thus the channel power gains are static within each time frame but may change across time frames. In order to obtain the upper bound of CE and provide theoretical support for the practical system design, It is assumed that the perfect CSI can be obtained \cite{F. Wang}, \cite{S. Bi}, \cite{Y. Mao1}-\cite{F. Zhou1}.

\begin{figure}[!t]
\centering
\includegraphics[width=3.0in]{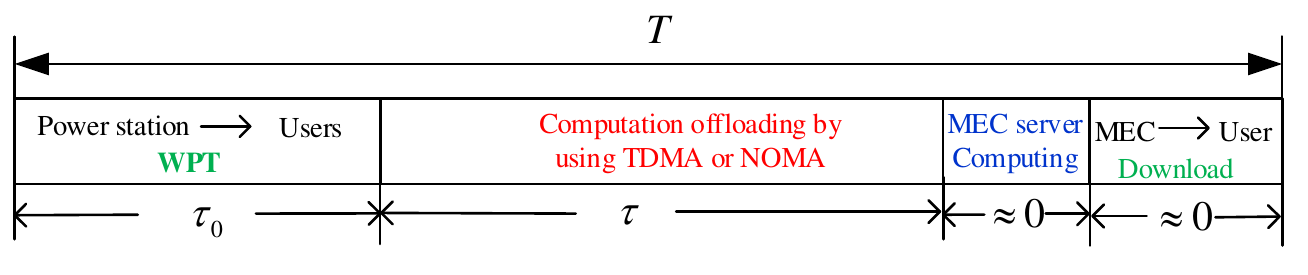}
\caption{The frame structure of the wireless powered MEC system.} \label{fig.1}
\end{figure}
The frame structure shown in Fig. 2 consists of four stages. The frame duration is denoted by $T$, which is selected based on the correlated time of the channel in order to guarantee that the channel power gains are constant within one frame duration \cite{F. Wang}. In the first stage, the wireless power station transfers energy to $K$ users. In the second stage, users offload their computation tasks to the MEC by using TDMA or NOMA protocol. In the third stage, the MEC executes the computation tasks from users. In the fourth stage, the MEC downloads the computation results to users. Similar to  \cite{C. You}-\cite{F. Zhou1}, the computation time and the downloading time of the MEC are neglected  as the MEC has a strong computation capability compared with users and the number of the bits related to the computation results is relatively small.

\begin{rem}
The major application scenarios  for wireless powered MEC networks include two cases.  One is in the wireless sensor or wearable networks where the mobile sensors and wearable computing devices are with milliwatt power consumption while they need to perform computation tasks, such as environmental parameter or physical condition  monitoring \cite{S. Bi}. The other one  is in the areas, such as wildernesses and complex terrains, where the government needs to keep monitoring the environment so that the corresponding strategies can be taken to protect the environment. In those areas, neither cable charging or battery replacement can be conveniently established nor the cost for establishing cable charging  systems is  affordable. The wireless powered MEC network becomes a desirable alternative \cite{C. You}.
\end{rem}
\subsection{Non-Linear Energy Harvesting Model}
Let $\tau_0$ denote the duration of the WPT stage. In this paper, different from the works in \cite{F. Wang}, \cite{S. Bi}, \cite{C. You}-\cite{F. Zhou1}, a practical non-linear EH model is applied while a sensitivity property is considered. Specifically, the harvested energy is zero when the input RF power is smaller than the sensitivity threshold. Based on the work in \cite{S. Wang}, the harvested energy of the $k$th user denoted by ${\Phi _k}\left( {{\tau _0},{P_s}} \right)$ can be given as
\begin{align}\label{27}\ \notag
&{\Phi _k}\left( {{\tau _0},{P_s}} \right)= \\
&{\tau _0}{\left[ {\frac{{P_k^{\max }}}{{\exp \left( { - \mu {P_0} + \psi } \right)}}\left( {\frac{{1 + \exp \left( { - \mu {P_0} + \psi } \right)}}{{1 + \exp \left( { - \mu {h_k}{P_s} + \psi } \right)}} - 1} \right)} \right]^{\rm{ + }}},
\end{align}
where $P_s$ is the transmit power of the power station; $P_k^{\max}$ is the maximum harvested power of the $k$th user, $k\in {\cal K}$ and ${\cal K}=1, 2, \cdots, K$; $P_0$ is the sensitivity threshold; $\mu$ and $\psi$ are the parameters for controlling the steepness of the function; $h_k$ is the instantaneous channel power gain from the power station to the $k$th user; ${\left[ a \right]^ + } = \max \left( {a,0} \right)$ and $\max \left( {a,0} \right)$ denotes the bigger value  of $a$ and $0$.

\begin{figure}[!t]
\centering
\includegraphics[width=3.0 in]{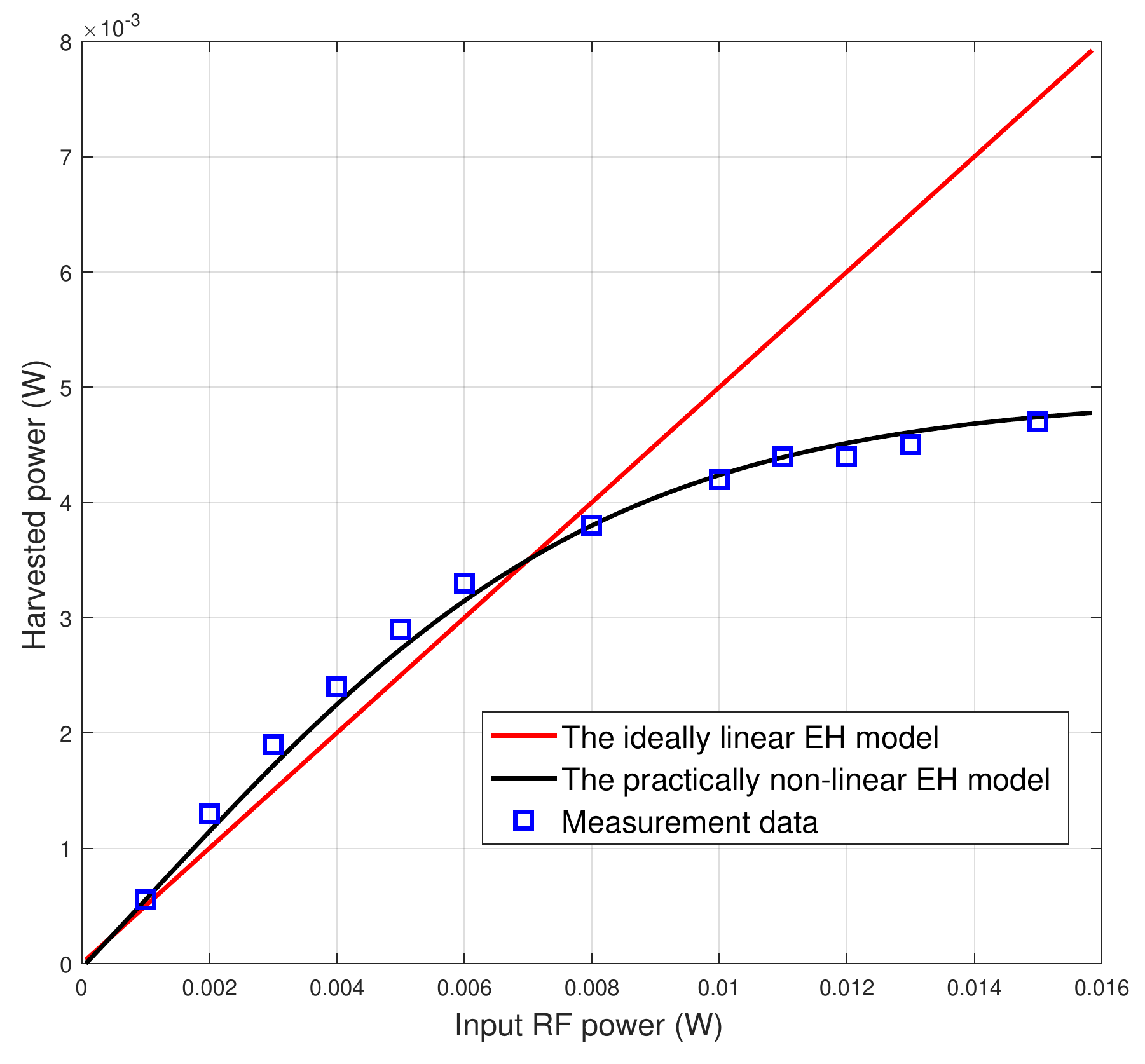}
\caption{A comparison of the harvested power based on the linear EH model, non-linear EH model and the measurement data \cite{S. Wang}.} \label{fig.1}
\end{figure}

In order to better illustrate  the non-linear EH model, Fig. 3 compares the harvested power obtained with the non-linear EH model \cite{S. Wang} to those achieved with the linear EH model and the experimental results given in \cite{S. Wang}. The harvested power of the linear EH model is $\varrho{h_k}{P_s}$, where $\varrho$ is the energy
conversion efficiency and selected as $0.5$. The parameters in the non-linear EH model are set as $P_k^{\max}=0.004927$ W, $P_0=0.000064$ W, $\mu=274$ and $\psi=0.29$.
\subsection{Partial Offloading}
In this mode, the computation tasks of each user can be divided into two parts,  one for local computing and  one for offloading.
\subsubsection{Local Computation} Similar to  \cite{F. Wang}, \cite{S. Bi}, \cite{C. You}-\cite{F. Zhou1}, each user can perform local computation in the entire frame as each user can have the circuit architecture to separate the computing unit and the offloading unit.  $C$ cycles are required for computing one bit of raw data at the user side CPU. Let $f_k$ denote the CPU frequency of the $k$th user. Thus, the number of locally computed bits  at the $k$th user and the consumed energy are $Tf_k/C$ and $T{\gamma _c}f_k^3$, respectively. $\gamma _c$ is the effective capacitance coefficient of the processor's chip, and $\gamma _c$  is dependent on the chip architecture.
\subsubsection{Offloading with TDMA}
Let $\tau_k$ and $P_k$ denote the offloading time and the transmit power for offloading of the $k$th user, respectively. Similar to the work in \cite{S. Bi}, the offloaded task of the $k$th user consists of raw data and communication overhead, such as the encryption and packet header. Let $v _k>1$ indicate the communication overhead. According to \cite{S. Bi},  the number of bits that the $k$th user offloads to the MEC server using TDMA is given as
$\frac{{B{\tau _k}}}{{{v_k}}}{\log _2}\left( {1 + \frac{{{g_k}{P_k}}}{{\sigma _0^2}}} \right), k\in {\cal K},$
where $B$ is the communication bandwidth, $\sigma _0^2$ denotes the noise power, and $g_k$ is the instantaneous channel power gain from the $k$th user to the MEC server. Thus, the CE of the $k$th user is defined as
\begin{subequations}
\begin{align}\label{27}\
&\eta_k\left( {\tau _0, {\tau _k},{P_k},{f_k}} \right)=\frac{{R_k\left( {{\tau _k},{P_k},{f_k}} \right)}}{{E_k\left( {\tau _0, {\tau _k},{P_k},{f_k}} \right)}},\\
&R_k\left( {{\tau _k},{P_k},{f_k}} \right) ={\frac{{T{f_k}}}{C} + \frac{{B{\tau _k}}}{{{v_k}}}{{\log }_2}\left( {1 + \frac{{{g_k}{P_k}}}{{\sigma _0^2}}} \right)},  \\
& E_k\left( {\tau _0, {\tau _k},{P_k},{f_k}} \right)={{{\tau _0}{P_{r,k}} + \zeta{\tau _k}\left({P_k} +{P_{c,k}}\right) + T{\gamma _c}f_k^3}},
\end{align}
\end{subequations}
where $\eta_k\left( {\tau _0, {\tau _k},{P_k},{f_k}} \right)$ is the CE of the $k$th user; $R_k\left( {{\tau _k},{P_k},{f_k}} \right)$ and $E_k\left( {\tau _0, {\tau _k},{P_k},{f_k}} \right)$ are the total number of computed bits at the MEC server for the $k$th user  and the consumption energy  of the $k$th user, respectively; $P_{r,k}$ and $P_{c,k}$ denote the received power for the received signal processing during the WPT stage and the constant circuit power consumption of the $k$th user during the computation offloading process, respectively \cite{Q. Wu2}, and $\zeta$ denotes the amplifier coefficient.

\subsubsection{Offloading with NOMA} Different from TDMA, NOMA enables all the users to simultaneously offload their offloading tasks on the same frequency band so that offloading throughput can be improved. Let $\tau_1$ denote the duration of the offloading process. Without loss of generality,  the channel gains for the NOMA users have a descending order ${g_1}<  \cdots  < {g_k} <\cdots  <  {g_K}$. Thus, using the simple decoding order based on the descending order of the channel power channel \cite{Z. Ding}-\cite{ S. Jeong}, the CE of the $k$th user can be expressed as
\begin{subequations}
\begin{align}\label{27}\
&\eta_k\left( {\tau _0, {\tau _1},{P_k},{f_k}} \right)=\frac{{R_k\left( {{\tau _1},{P_k},{f_k}} \right)}}{{E_k\left( {\tau _0, {\tau _1},{P_k},{f_k}} \right)}},\\
&{R_k}\left( {{\tau _1},{f_k},{P_k}} \right) = \left\{ \begin{array}{l}
\begin{array}{*{20}{c}}
{\frac{{T{f_k}}}{C} + \frac{{B{\tau _1}}}{{{v_k}}}{{\log }_2}\left( {1 + \frac{{{g_k}{P_k}}}{{\sum\limits_{i = k + 1}^K {{g_i}{P_i}}  + \sigma _0^2}}} \right),}&\\~~~~~~~~~~~~~~~{1 \le k \le K - 1},
\end{array}\\
\begin{array}{*{20}{c}}
{\frac{{T{f_k}}}{C} + \frac{{B{\tau _1}}}{{{v_k}}}{{\log }_2}\left( {1 + \frac{{{g_k}{P_k}}}{{\sigma _0^2}}} \right),}&{k = K},
\end{array}
\end{array} \right.
  \\
& E_k\left( {\tau _0, {\tau _1},{P_k},{f_k}} \right)={{{\tau _0}{P_{r,k}} + \zeta{\tau _1}\left({P_k} +{P_{c,k}}\right) + T{\gamma _c}f_k^3}},
\end{align}
\end{subequations}
where $\eta_k\left( {\tau _0, {\tau _1},{P_k},{f_k}} \right)$ denotes the CE of the $k$th user; ${R_k}\left( {{\tau _1},{f_k},{P_k}} \right)$ and $E_k\left( {\tau _0, {\tau _1},{P_k},{f_k}} \right)$ denote the total number of computed  bits  at MEC and the total energy consumption of the $k$th user respectively.
\subsection{Binary Offloading}
Under the binary offloading mode, the computation task can be either completely computed at the local device or completely offloaded to the MEC server for computing. Let ${\cal M}_1$ and ${\cal M}_0$ denote the set of users that choose to perform task offloading and the set of  users that choose to perform local computation, respectively. Thus, ${\cal K}={\cal M}_0\cup{\cal M}_1$ and ${\cal M}_0 \cap{\cal M}_1=\Theta$, where $\Theta$ denotes the null set.
\subsubsection{Local Computation}  In this case, all the harvested energy is used for local computing. Thus, the CE of the $m$th user denoted by ${\eta _m}\left( {{\tau _0},{f_m}} \right)$ can be given as
\begin{subequations}
\begin{align}\label{27}\
&{\eta _m}\left( {{\tau _0},{f_m}} \right){\rm{ = }}\frac{{\frac{{T{f_k}}}{C}}}{{{\tau _0}{P_{r,k}} + T{\gamma _c}f_k^3}}{\rm{ = }}\frac{{T{f_k}}}{{C\left( {{\tau _0}{P_{r,k}} + T{\gamma _c}f_k^3} \right)}},
  \\
&{R_m}\left( {{f_m}} \right){\rm{ = }}\frac{{T{f_m}}}{C}, \ m\in {\cal M}_0,\\
&{E_m}\left( {{\tau _0},{f_m}} \right){\rm{ = }}{\tau _0}{P_{r,m}} + T{\gamma _c}f_m^3,m\in {\cal M}_0,
\end{align}
\end{subequations}
where ${R_m}\left( {{f_m}} \right)$ and ${E_m}\left( {{\tau _0},{f_m}} \right)$ are the total number of locally computed bits and energy consumption for computation of the $m$th user respectively.
\subsubsection{Offloading with TDMA}  With TDMA for offloading, the computation efficiency of the $n$th user denoted by ${\eta _n}\left( {{\tau _0},{\tau _n},{P_n}} \right)$ can be given as
\begin{subequations}
\begin{align}\label{27}\
&{\eta _n}\left( {{\tau _0},{\tau _n},{P_n}} \right) = \frac{{{R_n}\left( {{\tau _n},{P_n}} \right)}}{{{E_n}\left( {{\tau _0},{\tau _n},{P_n}} \right)}}, \ n\in {\cal M}_1,
  \\
&{R_n}\left( {{\tau _n},{P_n}} \right) = \frac{{B{\tau _n}}}{{{v_n}}}{\log _2}\left( {1 + \frac{{{g_n}{P_n}}}{{\sigma _0^2}}} \right), \ n\in {\cal M}_1,\\
&{E_n}\left( {{\tau _0},{\tau _n},{P_n}} \right) = {\tau _0}{P_{r,n}} + \zeta {\tau _n}\left( {{P_n} + {P_{c,n}}} \right), \ n\in {\cal M}_1,
\end{align}
\end{subequations}
where ${R_n}\left( {{\tau _n},{P_n}} \right)$ and ${E_n}\left( {{\tau _0},{\tau _n},{P_n}} \right)$ are the number of computed bits and energy consumption for offloading of the $n$th user respectively.

\subsubsection{Offloading with NOMA} With NOMA,  the CE of the  $n$th user in ${\cal M}_1$ denoted by ${\eta _n}\left( {{\tau _0},{\tau _n},{P_n}} \right)$ can be given as
\begin{subequations}
\begin{align}\label{27}\
&{\eta _n}\left( {{\tau _0},{\tau _n},{P_n}} \right) = \frac{{{R_n}\left( {{\tau _n},{P_n}} \right)}}{{{E_n}\left( {{\tau _0},{\tau _n},{P_n}} \right)}},n\in {\cal M}_1,
  \\
&{R_n}\left( {{\tau _1},{P_n}} \right) = \frac{{B{\tau _1}}}{{{v_n}}}{\log _2}\left( {1 + \frac{{{g_n}{P_n}}}{{\sum\limits_{i > n,i \in {M_1}}^{} {{g_i}{P_i}} {\rm{ + }}\sigma _0^2}}} \right),\\
&{E_n}\left( {{\tau _0},{\tau _n},{P_n}} \right) = {\tau _0}{P_{r,n}} + \zeta {\tau _n}\left( {{P_n} + {P_{c,n}}} \right),n\in {\cal M}_1.
\end{align}
\end{subequations}

\section{CE Maximization In Wireless Powered MEC Networks: TDMA based}
\subsection{Partial Offloading Mode}
\subsubsection{Problem Formulation}
When the TDMA protocol and the partial computation offloading mode are applied, the CE maximization problem is formulated under the max-min fairness criterion as
\begin{subequations}
\begin{align}\label{27}\
&\mathbf{P}_{{1}}: {\mathop {\max }\limits_{{\tau _0},{\tau _k},{P_k},{P_s},{f_k}} }\ {\mathop {\min }\limits_{k \in{\cal K}}\ \eta\left( {\tau _0, {\tau _k},{P_k},{f_k}} \right)}\\
&\text{s.t.} \ \  C1:{\frac{{T{f_k}}}{C} + \frac{{B{\tau _k}}}{{{v_k}}}{{\log }_2}\left( {1 + \frac{{{g_k}{P_k}}}{{\sigma _0^2}}} \right)}\geq {R_{k,\min}},k\in {\cal K},\\
& C2:{\tau _0}{P_{r,k}} +  \zeta{\tau _k}\left({P_k} +{P_{c,k}}\right) + T{\gamma _c}f_k^3 \le {\Phi _k}\left( {{\tau _0},{P_s}} \right),\\
&  C3:\sum\limits_{n = 0}^K {{\tau _n}}  \le T,\\
&  C4: 0 \le {\tau _0} \le T, 0 \le {\tau _k} \le T, \ k\in {\cal K},\\
&  C5: 0\leq{P_s} \le {P_{th}}, {P_k} \ge 0, \ k\in {\cal K},\\
&  C6:  {f_k} \ge 0, \ k\in {\cal K}.
\end{align}
\end{subequations}
$R_{k,\min}$ is the minimum number of computed bits required by the $k$th user and $P_{th}$ is the maximum transmission power of the wireless power station. The constraint $C1$ is the minimum  computed bits constraint. The constraint $C2$ is the EH causal constraint that the total consumption energy cannot be larger than the harvested energy. The constraints $C3$ and $C4$ are the constraints on the EH time and the computation offloading time. $\mathbf{P}_{{1}}$ is a non-convex fractional optimization problem. It is challenging to solve $\mathbf{P}_{{1}}$ due to the existence of coupling relationship among different optimization variables, such as the coupling between $t_k$ and $P_k$, as well as due to  the non-convex constraints $C1$ and $C2$.
\subsubsection{Solution and Iterative Algorithm}
In order to solve $\mathbf{P}_{{1}}$, Theorem 1 is presented as follows.
\begin{myTheo}
In wireless powered MEC networks with TDMA under the partial computation offloading mode, the maximum CE under the max-min fairness criterion is achieved when $P_s=P_{th}$.
\end{myTheo}

\emph{Proof:} Let $\left\{\tau _0^{*}, {\tau _k^{*}}, {P_k^{*}}, {f_k^{*}}, {P_s^{*}}\right\}$ denote the optimal solution of $\mathbf{P}_{{1}}$. $\eta^{*}$ denotes the maximum CE achieved under the max-min fairness criterion.  It is not difficult to prove that $\eta^{*}\geq0$ and ${P_s^{*}}\geq P_0$. It is assumed that ${P_s^{*}}<P_{th}$. Let $\left\{\tau _0^{\ddag}, {\tau _k^{\ddag}}, {P_k^{\ddag}}, {f_k^{\ddag}}, {P_s^{\ddag}}\right\}$ denote another solution of $\mathbf{P}_{{1}}$ satisfying  ${P_s^{\ddag}}=P_{th}$,  ${\tau _k^{\ddag}}={\tau _k^{*}}$, ${P_k^{\ddag}}={P_k^{*}}$, ${f _k^{\ddag}}={f _k^{*}}$ and ${\Phi _k}\left( {{\tau _0^{\ddag}},{P_s^{\ddag}}} \right)={\Phi _k}\left( {{\tau _0^{*}},{P_s^{*}}} \right)$. $\eta^{\ddag}$ denotes the corresponding maximum CE. When ${P_s^{*}}$ is not large enough to achieve the maximum output power $P_k^{\max}$, one has ${\tau _0^{*}}>\tau _0^{\ddag}$ since ${P_s^{*}}<P_s^{\ddag}$.  It is quite evident that $\left\{\tau _0^{\ddag}, {\tau _k^{\ddag}}, {P_k^{\ddag}}, {f_k^{\ddag}}, {P_s^{\ddag}}\right\}$ satisfies all the constraints of $\mathbf{P}_{{1}}$. Since $E_k\left( {\tau _0^{\ddag}, {\tau _k^{\ddag}},{P_k^{\ddag}},{f_k^{\ddag}}} \right)<E_k\left( {\tau _0^{*}, {\tau _k}^{*},{P_k}^{*},{f_k}}^{*} \right)$ and $R_k\left( {{\tau _k^{\ddag}},{P_k^{\ddag}},{f_k^{\ddag}}} \right)=R_k\left( {{\tau _k^{*}},{P_k^{*}},{f_k^{*}}} \right)$, one has $\eta^{\ddag}>\eta^{*}$. This contradicts the assumption that $\left\{\tau _0^{*}, {\tau _k^{*}}, {P_k^{*}}, {f_k^{*}}, {P_s^{*}}\right\}$ is the optimal solution. Thus, ${P_s^{*}}=P_{th}$. When ${P_s^{*}}$ is  large to achieve the maximum output power $P_k^{\max}$, since ${\Phi _k}\left( {{\tau _0^{\ddag}},{P_s^{\ddag}}} \right)={\Phi _k}\left( {{\tau _0^{*}},{P_s^{*}}} \right)$, one has ${\tau _0^{*}}=\tau _0^{\ddag}$ and thus $\eta^{\ddag}=\eta^{*}$. Thus, $\left\{\tau _0^{\ddag}, {\tau _k^{\ddag}}, {P_k^{\ddag}}, {f_k^{\ddag}}, {P_s^{\ddag}}\right\}$ is also the optimal solution. Theorem 1 is proved.

\begin{rem}
It can be seen from Theorem 1 that the maximum CE achieved under the max-min fairness criterion increases with the transmission power of the power station in the wireless powered MEC networks with TDMA. If the transmission power level of the power station is not large enough for achieving the maximum harvested power of the user, the CE can be increased by increasing the transmission power of the power station.
\end{rem}

Motivated by the Dinkelbach's method \cite{Q. Wu}, Lemma 1 is given to transform $\mathbf{P}_{{1}}$ to a tractable problem.
\begin{lemma}
The optimal solution of $\mathbf{P}_{{1}}$ can be obtained if and only if the following equation holds.
\begin{subequations}
\begin{align}\label{27}\
&{\mathop {\max }\limits_{{\tau _0},{\tau _k},{P_k},{f_k}} }\ {\mathop {\min }\limits_{k \in{\cal K}}}\ R_k\left( {{\tau _k},{P_k},{f_k}} \right)-\eta^{*}E_k\left( {\tau _0, {\tau _k},{P_k},{f_k}} \right)\\
&= {\mathop {\min }\limits_{k \in{\cal K}}}\ R_k\left( {{\tau _k^{*}},{P_k^{*}},{f_k^{*}}} \right)-\eta^{*}E_k\left( {\tau _0^{*}, {\tau _k}^{*},{P_k}^{*},{f_k^{*}}} \right)=0,
\end{align}
\end{subequations}
\end{lemma}
where $\eta^{*}$ and $*$ denote the maximum CE and optimality, respectively.  The proof can be readily obtained from the generalized fractional programming theory \cite{G. M. N.Guerekata}.

Based on Lemma 1, $\mathbf{P}_{{1}}$ can be solved by solving a parameter problem, denoted by $\mathbf{P}_{{2}}$, given as
\begin{subequations}
\begin{align}\label{27}\
&\mathbf{P}_{{2}}: {\mathop {\max }\limits_{{\tau _0},{\tau _k},{P_k},{f_k}} }\ {\mathop {\min }\limits_{k \in{\cal K}}}\ R_k\left( {{\tau _k},{P_k},{f_k}} \right)-\eta E_k\left( {\tau _0, {\tau _k},{P_k},{f_k}} \right) \\
&\text{s.t.} \ \  C1-C6.
\end{align}
\end{subequations}
Here $\eta$ is a non-negative parameter. Although $\mathbf{P}_{{2}}$ is more tractable, it is still non-convex and has coupling among optimization variables. Auxiliary variables $y_k$ are further introduced, where $y_k=\tau{ _k}P_k, k\in {\cal K}$. Using the auxiliary variables and Theorem 1, $\mathbf{P}_{{2}}$ can be equivalently expressed as
\begin{subequations}
\begin{align}\label{27}\
&\mathbf{P}_{{3}}: {\mathop {\max }\limits_{{\tau _0},{\tau _k},{y_k},{f_k},\Upsilon} }\ \Upsilon\\
&\text{s.t.}\ \frac{{T{f_k}}}{C} + \frac{{B{\tau _k}}}{{{v_k}}}{\log _2}\left( {1 + \frac{{{g_k}{y_k}}}{{{\tau _k}\sigma _0^2}}} \right) \ge {R_{k,\min}},k\in {\cal K},\\
&{\tau _0}{P_{r,k}} +  \zeta{y_k} + \zeta{\tau _k}{P_{c,k}} + T{\gamma _c}f_k^3 \le {\Phi _k}\left( {{\tau _0},{P_{th}}} \right),\\ \notag
&\ \ \ \ \ \ \frac{{T{f_k}}}{C} + \frac{{B{\tau _k}}}{{{v_k}}}{\log _2}\left( {1 + \frac{{{g_k}{y_k}}}{{{\tau _k}\sigma _0^2}}} \right)-\\
& \eta \left[ {{\tau _0}{P_{r,k}} + \zeta{y_k} + \zeta{\tau _k}{P_{c,k}} + T{\gamma _c}f_k^3} \right] \ge   \Upsilon, k\in {\cal K},\\
& C3, C4, C6, y_k\geq0, k\in {\cal K}.
\end{align}
\end{subequations}

 \begin{lemma}
 $\mathbf{P}_{{3}}$ is convex and can be efficiently solved by using the  convex optimization tool  \cite{S. P. Boyd}.
\end{lemma}

\emph{Proof:} Firstly, it is evident that the objective function and the constraints $C3$, $C4$, $C6$ of $\mathbf{P}_{{3}}$ satisfy the conditions of a convex problem since the objective function is linear and the constraints $C3$, $C4$, $C6$ are linear inequality constraints. For the constraint given by  $\left(11\rm{b}\right)$, $\frac{{T{f_k}}}{C}$ is a linear function with respect to $f_k$  and $\frac{{B{\tau _k}}}{{{v_k}}}{\log _2}\left( {1 + \frac{{{g_k}{y_k}}}{{{\tau _k}\sigma _0^2}}} \right)$ is  the perspective of $\frac{{B}}{{{v_k}}}{\log _2}\left( {1 + \frac{{{g_k}{y_k}}}{{\sigma _0^2}}} \right)$, which is a concave function of $y_k$. Since the perspective operation preserves convexity \cite{S. P. Boyd}, $\frac{{B{\tau _k}}}{{{v_k}}}{\log _2}\left( {1 + \frac{{{g_k}{y_k}}}{{{\tau _k}\sigma _0^2}}} \right)$ is concave with respect to $\tau _k$ and $y_k$. Thus, it is easy to obtain that the constraint given by $\left(11\rm{b}\right)$ is a convex constraint. For the constraint ${\tau _0}{P_{r,k}} +  \zeta{y_k} + \zeta{\tau _k}{P_{c,k}} + T{\gamma _c}f_k^3 \le {\Phi _k}\left( {{\tau _0},{P_{th}}} \right)$, the right side  ${\Phi _k}\left( {{\tau _0},{P_{th}}} \right)$ is a linear function with respect to $\tau _0$ and the left side ${\tau _0}{P_{r,k}} +  \zeta{y_k} + \zeta{\tau _k}{P_{c,k}} + T{\gamma _c}f_k^3$ is a linear function in regard to ${\tau _0}$, $y_k$ and $\tau _k$. Moreover, since the local CPU frequency $f_k$ is nonnegative, ${\tau _0}{P_{r,k}} +  \zeta{y_k} + \zeta{\tau _k}{P_{c,k}} + T{\gamma _c}f_k^3$ is a convex function with respect to $f_k$ when $f_k\geq 0$. Thus, the constraint given by $\left(10\rm{c}\right)$ is also a convex constraint. Using the same analysis method for the constraint given by $\left(10\rm{d}\right)$, it is easy to prove that  the constraint given by $\left(10\rm{d}\right)$ is also a convex constraint. Thus, it is proved that $\mathbf{P}_{{3}}$ is convex.

In this paper, in order to gain more meaningful insights, the optimal solutions are obtained in closed forms by using the Lagrange duality method \cite{G. M. N.Guerekata}. Towards that end, let $f _k^{*}$ and $P _k^{*}$ denote the optimal local computation frequency and the optimal offloading power of the $k$th user, $k\in {\cal K}$, respectively. By solving $\mathbf{P}_{{3}}$, Theorem 2 can be stated as follows.
\begin{myTheo}
In the wireless powered MEC systems with TDMA, the optimal local computation frequency $f_k^{*}$ and the optimal offloading power $P_k^{*}$ of the $k$th user for maximizing the CE under the max-min fairness criterion have the following mathematical expressions:
\begin{subequations}
\begin{align}\label{27}\
&f_k^{*} = \sqrt {\frac{{{\lambda _k} + {\theta _k}}}{{3C{\gamma _c}\left( {{\rho _k} + {\theta _k}\eta } \right)}}};\\
&P_k^{*} = \left\{ \begin{array}{l}
\begin{array}{*{20}{c}}
{0,}&{\text{if}\ {\tau _k} = 0}
\end{array}\\
\begin{array}{*{20}{c}}
{{{\left[ {\frac{{\left( {{\lambda _k} + {\theta _k}} \right)B}}{{\zeta{v_k}\ln 2\left( {{\rho _k} + {\theta _k}\eta } \right)}} - \frac{{\sigma _0^2}}{{{g_k}}}} \right]}^ + },}&{\text{otherwise}};
\end{array}
\end{array} \right.
\end{align}
\end{subequations}
where $\lambda _k\geq0$, $\rho _k\geq0$ and $\theta _k\geq0$ are the dual variables corresponding to the constraints given by $\left(7\rm{b}\right)$, $\left(7\rm{c}\right)$, and $\left(7\rm{d}\right)$,  respectively.
\end{myTheo}
\begin{IEEEproof}
See Appendix A.
\end{IEEEproof}
\begin{rem}
It can be seen from Theorem 2 that the $k$th user chooses to offload its computation task only when the channel between the $k$th user and the MEC server is good enough, namely, $g_k\geq {\left[ {\sigma _0^2\zeta{v_k}\ln 2\left( {{\rho _k} + {\theta _k}\eta } \right)} \right]}/{\left( {{\lambda _k} + {\theta _k}} \right)B}$. Moreover, the  local computation frequency decreases with the increase of  $\eta$, which is related to the CE. It indicates that the users prefer to offload their computation task to the MEC server in order to improve the CE.  Furthermore, it can be seen that the users prefer to offload their computation task when the local CPU frequency is too high, namely, $f_k^{*} \ge \sqrt {\frac{{\sigma _0^2\zeta{v_k}\ln 2}}{{{g_k}3C{\gamma _c}B}}}$. Additionally, when $f_k^{*} < \sqrt {\frac{{\sigma _0^2\zeta{v_k}\ln 2}}{{{g_k}3C{\gamma _c}B}}}$, the optimal offloading power $P_k^{*}$ is zero. It means that the users only perform local computation in order to maximize their CE.
\end{rem}

By solving $\mathbf{P}_{{3}}$, Theorem 3 is stated to clarify the characteristic of the EH time $\tau_0$ and  $\tau_k$.
\begin{myTheo}
For  the given $\lambda _k$, $\rho _k$, $\beta$ and $\theta _k$, in order to maximize the Lagrangian of $\mathbf{P}_{{3}}$, the optimal EH time $\tau_0^{*}$ and computation offloading time $\tau_k^{*}$ need to satisfy the following equations.
\begin{subequations}
\begin{align}\label{27}\
&\tau _0^{*} = \left\{ \begin{array}{l}
\begin{array}{*{20}{c}}
{0,}&{\text{if}\ z\left( {{\rho _k},\beta ,{\theta _k}} \right) < 0},
\end{array}\\
\begin{array}{*{20}{c}}
{ \in \left[ {0,T} \right),}&{\text{if} \ z\left( {{\rho _k},\beta ,{\theta _k}} \right) = 0},
\end{array}
\end{array} \right.
\\
&z\left( {{\rho _k},\beta ,{\theta _k}} \right) = \sum\limits_{k = 1}^K {{\rho _k}} \left( {{P_{E,k}} - {P_{r,k}}} \right) - \beta  - \sum\limits_{k = 1}^K {{\theta _k}\eta } {P_{r,k}},\\
&{P_{E,k}} ={\frac{{P_k^{\max }}}{{\exp \left( { - \mu {P_0} + \psi } \right)}}\left( {\frac{{1 + \exp \left( { - \mu {P_0} + \psi } \right)}}{{1 + \exp \left( { - \mu {h_k}{P_{th}} + \psi } \right)}} - 1} \right)},
\end{align}
\end{subequations}
\begin{subequations}
\begin{align}\label{27}\
&\tau _k^{*} = \left\{ \begin{array}{l}
\begin{array}{*{20}{c}}
{{\rm Z} ,}&{\text{if} \ \frac{{\sigma _0^2\zeta{v_k}{\theta _k}\eta \ln 2}}{{\left( {{\lambda _k} + {\theta _k}} \right)B}} > \omega^{opt} },
\end{array}\\
\begin{array}{*{20}{c}}
{ \in \left[ {0,{\rm Z} } \right],}&{\text{if} \ \frac{{\sigma _0^2\zeta{v_k}{\theta _k}\eta \ln 2}}{{\left( {{\lambda _k} + {\theta _k}} \right)B}} = \omega^{opt} },
\end{array}\\
\begin{array}{*{20}{c}}
{0,}&{\text{if} \ \frac{{\sigma _0^2\zeta{v_k}{\theta _k}\eta \ln 2}}{{\left( {{\lambda _k} + {\theta _k}} \right)B}} < \omega^{opt} },
\end{array}
\end{array} \right.
\\
&{\rm Z } = \frac{{{\Phi _k}\left( {\tau _0^{opt},{P_{th}}} \right) - \tau _0^{opt}{P_{r,k}} - T{\gamma _c}{{\left( {f_k^{opt}} \right)}^3}}}{{{P_{c,k}} + P_k^{opt}}},
\end{align}
\end{subequations}
where $\omega^{*} $ is the solution of the following equation.
\begin{align}\label{27}\ \notag
&\frac{{B\left( {{\lambda _k} + {\theta _k}} \right)}}{{{v_k}}}{\log _2}\left[ {\frac{{\left( {{\lambda _k} + {\theta _k}} \right)Bw}}{{\zeta{v_k}{\theta _k}\eta \sigma _0^2\ln 2}}} \right] - \frac{{\left( {{\lambda _k} + {\theta _k}} \right)B}}{{{v_k}\ln 2}}\\
&+ \frac{{\zeta\left( {{\rho _k} + {\theta _k}\eta } \right) \sigma _0^2}}{w} - \zeta\left( {{\rho _k} + {\theta _k}\eta } \right) {P_{c,k}} - \beta  = 0.
\end{align}
\end{myTheo}
\begin{IEEEproof}
See Appendix B.
\end{IEEEproof}
By solving $\mathbf{P}_{{3}}$, Theorem 4 can be stated to obtain the maximum CE denoted by $\Upsilon^{*}$.
\begin{myTheo}
In the wireless powered MEC systems with TDMA under the partial computation offloading mode, the maximum CE under the max-min fairness criterion is given as
\begin{subequations}
\begin{align}\label{27}\
&\Upsilon ^{*}= \left\{ \begin{array}{l}
\begin{array}{*{20}{c}}
{0,}&{\text{if}\ \sum\limits_{k = 1}^K { {\theta _k}}  > 1,}
\end{array}\\
\begin{array}{*{20}{c}}
{{\Lambda ^*,}}&{\text{if}\ \sum\limits_{k = 1}^K { {\theta _k}}  \le 1;}
\end{array}
\end{array} \right.
\\
&\Lambda ^*={\mathop {\min }\limits_{k \in{\cal K}}}\ R_k\left( {{\tau _k^{*}},{P_k^{*}},{f_k^{*}}} \right)-\eta E_k\left( {\tau _0^{*}, {\tau _k}^{*},{P_k}^{*},{f_k^{*}}} \right),
\end{align}
\end{subequations}
\end{myTheo}
\begin{IEEEproof}
Since $\mathbf{P}_{{3}}$ is convex and the Slater's conditions are satisfied, the Lagrangian of $\mathbf{P}_{{3}}$ should be upper-bound with respect to $\Upsilon$. When $\sum\limits_{k = 1}^K { {\theta _k}}  > 1$, the Lagrangian decreases with $\Upsilon$. Thus, the maximum of Lagrangian is achieved with $\Upsilon=0$ since $\Upsilon\geq0$. When  $\sum\limits_{k = 1}^K { {\theta _k}} \leq 1$, the maximum of Lagrangian is obtained when the optimal solution is achieved.
\end{IEEEproof}

Finally, an iterative algorithm denoted by Algorithm 1 is given to obtain the maximum CE. Specifically, when $\left| \Upsilon ^{*,n}-\eta _{}^{n}\right|=0$, the optimal solution is obtained, where $n$ and $\Upsilon ^{*,n}$ denote the iterative number and the optimal solution achieved at the $n$th iteration, respectively. Otherwise, an $\xi$-optimal solution is adopted with an error tolerance $\xi$. In other words, the maximum CE is obtained when $\left| \Upsilon ^{*,n}-\eta _{}^{n}\right| \le \xi$, where $\left| \cdot \right|$ denotes the absolute operator. The details of Algorithm 1 are given in Table I.

\begin{rem}
By using Theorem 1 and introducing auxiliary variables $y_k$, it is seen that  $\mathbf{P}_{{1}}$ is a generalized fractional programming problem \cite{G. M. N.Guerekata}. Moreover, Algorithm 1 is proposed for solving $\mathbf{P}_{{1}}$ based on the Dinkelbach's method. Thus, according to \cite{G. M. N.Guerekata}, Algorithm 1 can converge when updating $\eta _{}^{n}$. The detail proof for the convergence can be seen in \cite{G. M. N.Guerekata}.
\end{rem}

\begin{table}[htbp]
\begin{center}
\caption{The iterative algorithm}
\begin{tabular}{lcl}
\\\toprule
$\textbf{Algorithm 1}$: The iterative algorithm for $\mathbf{P}_{\text{1}}$\\ \midrule
\  1) \textbf{Input settings:}\\
 \ \ \ \ \ \ \ the error tolerance $\xi> 0$, $R_{k,\min} > 0$ and $P_{th}$,\\
  \ \ \ \ \ \ \ the maximum iteration number $N$.\\
\  2) \textbf{Initialization:}\\
 \ \ \ \ \ \ \ EE $\eta _{}^{n} =\eta_0$ and the iteration index $n=0$.\\
\  3) \textbf{Optimization:}\\
\ \ \ \ \ \textbf{$\pmb{\unrhd} $  for \emph{n}=1:N }\\
\ \ \ \ \ \ \ \ \ \ solve $\text{P}_{\textbf{3}}$ by using \texttt{CVX} for the given $\eta _{}^{n}$;\\
\ \ \ \ \ \ \ \ \ \ obtain the solution $\left\{{\tau _0^{*,n}, {\tau _k}^{*,n},{P_k}^{*,n},{f_k^{*,n}},\Upsilon ^{*,n}}\right\}$;\\
\ \ \ \ \ \ \ \ \ \  \textbf{if} $\left| \Upsilon ^{*,n}-\eta _{}^{n}\right| \le \xi$ \\
\ \ \ \ \ \ \ \ \ \ \ \ \ \ the maximum CE $\Upsilon^{*,n}$ is obtained;\\
\ \ \ \ \ \ \ \ \ \ \ \ \ \ break;\\
\ \ \ \ \ \ \ \ \ \  \textbf{else} \\
\ \ \ \ \ \ \ \ \ \ \ \ \ \  update $n=n+1$ and $\eta _{}^{n}=\Upsilon ^{*,n}$;\\
\ \ \ \ \ \ \ \ \ \  \textbf{end}\\
\ \ \ \ \ \textbf{$\pmb{\unrhd} $ end} \\
\bottomrule
\end{tabular}
\end{center}
\end{table}
\subsection{Binary Offloading Mode}
\subsubsection{Problem Formulation}Under the binary  offloading mode, when the TDMA protocol is applied, the CE maximization problem under the max-min fairness criterion can be formulated as $\mathbf{P}_{{4}}$.
\begin{subequations}
\begin{align}\label{27}\
&{\mathop {\max }\limits_{{\tau _0},{\tau _n},{P_n},{P_s},{f_m}} }\ {\mathop {\min }\limits_{m \in {M_0},n \in {M_1}} }
\left\{ {{\eta _m}\left( {{\tau _0},{f_m}} \right),{\eta _n}\left( {{\tau _0},{\tau _n},{P_n}} \right)} \right\}
\\
&\text{s.t.} \ \ \frac{{T{f_m}}}{C} \geq {R_{m,\min}},m\in {\cal M}_0,\\
&\frac{{B{\tau _n}}}{{{v_n}}}{\log _2}\left( {1 + \frac{{{g_n}{P_n}}}{{\sigma _0^2}}} \right)\geq {R_{n,\min}},n\in {\cal M}_1,\\
&{\tau _0}{P_{r,m}} + T{\gamma _c}f_m^3 \le {\Phi _m}\left( {{\tau _0},{P_s}} \right),m \in {\cal M}_0\\
&{\tau _0}{P_{r,n}}  + \zeta {\tau _n}\left( {{P_n} + {P_{c,n}}} \right) \le {\Phi _n}\left( {{\tau _0},{P_s}} \right),n \in {M_1}\\
&\sum\limits_{n \in {{\cal M}_1}} {{\tau _n}}  \le T, 0 \le {\tau _n} \le T,\\
& {P_n} \ge 0,n \in {\cal M}_1, {f_m} \ge 0, m \in {\cal M}_0, \ \text{and} \ 0\leq{P_s} \le {P_{th}},
\end{align}
\end{subequations}
where the constraints given by $\left(16\rm{b}\right)$ and $\left(16\rm{c}\right)$ are the requirements of the minimum computed bits. The constraints given by $\left(16\rm{d}\right)$ and $\left(16\rm{e}\right)$ are the EH causal constraints. $\mathbf{P}_{{4}}$ is challenging to solve. Moreover, it is impractical to use the exhaustive search method for determining the operational mode selection due to the extremely high complexity, especially when the number of users is large.
\subsubsection{Alternative Optimization Algorithm} In order to solve $\mathbf{P}_{{4}}$,  let ${{\alpha _k}}=0$ indicate that the $k$th user performs local computation mode and ${{\alpha _k}}=1$ mean that the $k$th user  performs task offloading, where $k\in {{\cal K}}$. Moreover, $\alpha _k$ is relaxed as a continuous sharing factor $\alpha _k\in \left[0, 1\right]$. Thus, $\mathbf{P}_{{4}}$ can be expressed as
\begin{subequations}
\begin{align}\label{27}\ \notag
&\mathbf{P}_{{5}}: {\mathop {\max }\limits_{{\tau _0},{\tau _k},{P_k},{P_s},{f_k},\alpha_k} }\ {\mathop {\min }\limits_{k \in {{\cal K}}} }\ \
{\eta _k}\left( {{\tau _0},{\tau _k},{P_k},{f_k},{\alpha _k}} \right) = \\
&\frac{{\left( {1 - {\alpha _k}} \right)\frac{{T{f_k}}}{C} + {\alpha _k}\frac{{B{\tau _k}}}{{{v_k}}}{{\log }_2}\left( {1 + \frac{{{g_k}{P_k}}}{{\sigma _0^2}}} \right)}}{{\left( {1 - {\alpha _k}} \right)\left( {{\tau _0}{P_{r,k}} + T{\gamma _c}f_k^3} \right) + {\alpha _k}\left[ {{\tau _0}{P_{r,k}} + {\tau _k}\zeta \left( {{P_k} + {P_{c,k}}} \right)} \right]}}
\\
&\text{s.t.} \ \ \left( {1 - {\alpha _k}} \right)\frac{{T{f_k}}}{C} + {\alpha _k}\frac{{B{\tau _k}}}{{{v_k}}}{\log _2}\left( {1 + \frac{{{g_k}{P_k}}}{{\sigma _0^2}}} \right) \geq {R_{k,\min}}, \\ \notag
&\left( {1 - {\alpha _k}} \right)\left( {{\tau _0}{P_{r,k}} + T{\gamma _c}f_k^3} \right) + {\alpha _k}\left[ {{\tau _0}{P_{r,k}} + {\tau _k}\zeta \left( {{P_k} + {P_{c,k}}} \right)} \right]\\
&\le {\Phi _k}\left( {{\tau _0},{P_s}} \right), \ k \in {\cal K},\\
&{\tau _0}{\rm{ + }}\sum\limits_{k = 1}^K {{\alpha _k}{\tau _k}}  \le T, C4-C6, \ 0 \le {\alpha _k} \le 1.
\end{align}
\end{subequations}
It can be seen from $\left(17\right)$ that $\mathbf{P}_{{5}}$ is similar to $\mathbf{P}_{{1}}$. Thus, for a given $\alpha _k$, the method for solving $\mathbf{P}_{{1}}$ can be applied to solve $\mathbf{P}_{{5}}$. Moreover, it is easy to prove that $\mathbf{P}_{{5}}$ is a linear fractional optimization problem when other optimization variables are given. Thus, an alternative optimization algorithm denoted by Algorithm 2 is proposed. In order to tackle the non-convexity of the constraint given by $\left(17\rm{c}\right)$, Lemma 2 is given.
\begin{lemma}
In wireless powered MEC systems with TDMA under the binary offloading mode, the maximum CE under the max-min fairness criterion is achieved when $P_s=P_{th}$.
\end{lemma}

Using Lemma $3$ and the same method for solving $\mathbf{P}_{{1}}$, for given $\alpha _k$, $\text{P}_{{5}}$ can be solved by solving the following problem $\mathbf{{P}_6}$.
\begin{subequations}
\begin{align}\label{27}\
&\mathbf{P}_{{6}}: {\mathop {\max }\limits_{{\tau _0},{\tau _k},{y_k},{f_k},\Upsilon} }\ \Upsilon\\
&\text{s.t.}\ \ \left( {1 - {\alpha _k}} \right) \frac{{T{f_k}}}{C} + {\alpha _k}\frac{{B{\tau _k}}}{{{v_k}}}{\log _2}\left( {1 + \frac{{{g_k}{y_k}}}{{{\tau _k}\sigma _0^2}}} \right) \ge {R_{k,\min}}, \\ \notag
&{\tau _0}{P_{r,k}} +  {\alpha _k}\zeta{y_k} + {\alpha _k}\zeta{\tau _k}{P_{c,k}} + \left( {1 - {\alpha _k}} \right)T{\gamma _c}f_k^3 \\
&\le {\Phi _k}\left( {{\tau _0},{P_{th}}} \right),\\ \notag
&\left( {1 - {\alpha _k}} \right)\frac{{T{f_k}}}{C} + {\alpha _k}\frac{{B{\tau _k}}}{{{v_k}}}{\log _2}\left( {1 + \frac{{{g_k}{y_k}}}{{{\tau _k}\sigma _0^2}}} \right)\\
&- \eta \left[ {{\tau _0}{P_{r,k}} + {\alpha _k}\zeta{y_k} +{\alpha _k} \zeta{\tau _k}{P_{c,k}} + \left( {1 - {\alpha _k}} \right)T{\gamma _c}f_k^3} \right] \ge   \Upsilon\\
& C4, C6, \left(17\rm{d}\right), y_k\geq0,  k\in {\cal K}.
\end{align}
\end{subequations}
In the above $y_k={\tau _k}P_k$; $\Upsilon\geq0$ and $\eta\geq0$ are auxiliary variables. Moreover, it can be seen that  $\mathbf{P}_{{6}}$ is convex in terms of ${\alpha _k}$ when other optimization variables are given. Thus, using the Lagrange duality method, the operational mode selection variables ${\alpha _k}$ can be obtained by using the following Theorem.
\begin{myTheo}
In the wireless powered MEC systems with TDMA under the binary computation offloading mode, the optimal operational mode selection index has the following form
\begin{subequations}
\begin{align}\label{27}\
&{\alpha _k}^{*}= \left\{ \begin{array}{l}
0,\text{if}\ {F_{1,k}} < {F_{2,k}},\\
1,\text{otherwise};
\end{array} \right.
\\ \notag
&{F_{1,k}} = \left( {{\lambda _{{k}}} + {\chi  _k}} \right)\frac{{B{\tau _k}}}{{{v_k}}}{\log _2}\left( {1 + \frac{{{g_k}{{{y}}_k}}}{{{\tau _k}\sigma _0^2}}} \right)\\
 &\ \ \ \ \ \ \ - \zeta \left( {{\mu _k} + {\chi  _k}\eta } \right)\left( {{{{y}}_k} + {\tau _k}{P_{c,k}}} \right) - \upsilon {\tau _k},\\
&{F_{2,k}} = \left( {{\lambda _{{k}}} + {\chi  _k}} \right)\frac{{T{f_k}}}{C} - \left( {{\mu _k} + {\chi  _k}\eta } \right)T{\gamma _c}f_k^3,
\end{align}
\end{subequations}
where $\lambda _{{k}}\geq0$,  $\mu _k\geq0$, $\upsilon\geq0$ and $\chi  _k\geq0$ are the dual variables associated with the constraints given by $\left(18\rm{b}\right)$, $\left(18\rm{c}\right)$, $\left(18\rm{d}\right)$ and $\left(17\rm{d}\right)$.
\end{myTheo}
\begin{IEEEproof}
Theorem 5 can be readily proved from the derivations of the Lagrangian with respect to ${\alpha _k}$. The proof is omitted due to the space limit.
\end{IEEEproof}

It can be seen from Theorem 5 that the selection of the operation mode under the binary computation offloading mode depends on the tradeoff between the achievable computed  bits and the cost. Specifically, when this tradeoff under the local computing mode is better than that obtained under the complete offloading mode, the user chooses to perform local computing, and vice versa. Based on Algorithm 1 and Theorem 5, Algorithm 2 for solving $\mathbf{P}_{\text{4}}$ is presented in Table 2.

{\begin{rem}
The proof for the convergence of Algorithm 2 when updating $\alpha_k$ can be obtained from two facts. One is that $\mathbf{P}_{{4}}$ is a generalized fractional programming problem, which can be solved by using Algorithm 2 based on the Dinkelbach's method. Algorithm 2 can converge when iteratively updating $\eta$ \cite{G. M. N.Guerekata}. This indicates that the objective function  of $\mathbf{P}_{{6}}$ is nondecreasing, which can be easily proved by using the property of the generalized fractional programming problem \cite{G. M. N.Guerekata}.  The other is that $\mathbf{P}_{{6}}$ is convex in terms of $\alpha_k$. Thus, in each iteration, there is only an optimal solution of $\alpha_k$. Due to those two facts, it is easy to prove that  Algorithm 2 is converged when updating $\alpha_k$.
\end{rem}
\begin{table}[htbp]
\begin{center}
\caption{The alternative algorithm}
\begin{tabular}{lcl}
\\\toprule
$\textbf{Algorithm 3}$: The alternative algorithm for $\mathbf{P}_4$\\ \midrule
\  1) \textbf{Input settings:}\\
 \ \ \ \ \ \ \ the error tolerance $\xi_1> 0$, $\xi_2> 0$, $R_{k,\min} > 0$ and $P_{th}$,\\
  \ \ \ \ \ \ \ the maximum iteration number $N$.\\
\  2) \textbf{Initialization:}\\
 \ \ \ \ \ \ \ $\eta _{}^{n} =\eta_0$  and the iteration index $n=0$.\\
\  3) \textbf{Optimization:}\\
\ \ \ \ \ \textbf{$\pmb{\unrhd} $  for \emph{n}=1:N }\\
\ \ \ \ \ \ \ \ \ \ initialize the iteration index $j=1$ and ${\alpha_k^{j}}={\alpha_k^{1}}$;\\
\ \ \ \ \ \ \ \ \ \ \textbf{Repeat:}\\
\ \ \ \ \ \ \ \ \ \ \ \ \ \ solve $\text{P}_{\textbf{6}}$ by using \texttt{CVX} for the given $\eta _{}^{n}$ and ${\alpha_k^{j}}$ ;\\
\ \ \ \ \ \ \ \ \ \ \ \ \ \ obtain the solution $\left\{{\tau _{0,j}^{*,n}, {\tau _{k,j}^{*,n}},{P_{k,j}^{*,n}},{f_{k,j}^{*,n}},\Upsilon_j ^{*,n}}\right\}$;\\
\ \ \ \ \ \ \ \ \ \ \ \ \ \ use the subgradient method to update the dual variables;\\
\ \ \ \ \ \ \ \ \ \ \ \ \ \ update $j=j+1$   and ${\alpha_k^{j}}$;\\
\ \ \ \ \ \ \ \ \ \ \ \  \ \ \textbf{if} $\left| \Upsilon_j ^{*,n}-\Upsilon_{j-1} ^{*,n}\right| \le \xi_1$ \\
\ \ \ \ \ \ \ \ \ \ \ \ \ \ \ \ \ \ break;\\
\ \ \ \ \ \ \ \ \ \ \ \ \ \ \textbf{end}\\
\ \ \ \ \ \ \ \ \ \  \textbf{end Repeat} \\
\ \ \ \ \ \ \ \ \ \  \textbf{if} $\left| \Upsilon_j ^{*,n}-\eta _{}^{n}\right| \le \xi_2$ \\
\ \ \ \ \ \ \ \ \ \ \ \ \ \ the maximum CE $\Upsilon_j ^{*,n}$ is obtained;\\
\ \ \ \ \ \ \ \ \ \ \ \ \ \ break;\\
\ \ \ \ \ \ \ \ \ \  \textbf{else} \\
\ \ \ \ \ \ \ \ \ \ \ \ \ \  update $n=n+1$ and $\eta _{}^{n}=\Upsilon_j ^{*,n}$;\\
\ \ \ \ \ \ \ \ \ \  \textbf{end}\\
\ \ \ \ \ \textbf{$\pmb{\unrhd} $ end} \\
\bottomrule
\end{tabular}
\end{center}
\end{table}
\section{CE Maximization in Wireless Powered MEC Networks: NOMA Based}
In this section, CE maximization problems are studied in the wireless powered MEC networks with NOMA under both partial and binary computation offloading modes. The CE achieved under the max-min fairness criterion is maximized by jointly optimizing the CPU frequency, the EH time, the offloading power and time of users.  In order to tackle those non-convex optimization problems, an iterative algorithm and an alternative optimization algorithm based on SCA are proposed for solving the CE maximization problem under the partial and binary computation offloading mode, respectively.
\subsection{Partial Offloading Mode}
\subsubsection{Problem Formulation} When the NOMA protocol and the partial  computation offloading mode are considered, the CE maximization problem is formulated under the max-min fairness criterion as
\begin{subequations}
\begin{align}\label{27}\
&\mathbf{P}_{{7}}: {\mathop {\max }\limits_{{\tau _0},{\tau _1},{P_k},{P_s},{f_k}} }\ {\mathop {\min }\limits_{k \in{\cal K}}\ \eta_k\left( {\tau _0, {\tau _1},{P_k},{f_k}} \right)}\\
&\text{s.t.} \ \  {R_k}\left( {{\tau _1},{f_k},{P_k}} \right) \geq {R_{k,\min}}, \ k\in {\cal K},\\
&  \ \ \ \ \sum\limits_{i = 0}^1 {{\tau _i}}  \le T,\\
& \ \ \ \  0 \le {\tau _0} \le T, 0 \le {\tau _1} \le T,\\
&\ \ \ \  C2, C5 \ \text{and} \ C6.
\end{align}
\end{subequations}
$\mathbf{P}_{{7}}$ is challenging to tackle due to the minimum computation bit constraint given by $\left(21\rm{b}\right)$ and the non-convex constraint $C2$. In order to tackle it, an iteration algorithm based on SCA is proposed.
\subsubsection{Solution and The Iterative Algorithm}
In order to tackle the constraint $C2$, Lemma 4 is given.
\begin{lemma}
In wireless powered MEC systems with NOMA under the partial computation offloading mode, the maximum CE under the max-min fairness criterion is always achieved when $P_s=P_{th}$.
\end{lemma}

It is easy to prove that ${P_k}$ and $\tau _1$ are larger than zero, where $k\in {\cal K}$. Thus, in order to solve $\mathbf{P}_{{7}}$, auxiliary variables $x_k$ and $d_i$ are introduced, where ${P_k} = \exp \left( {{x_k}} \right)$ and ${\tau _1} = \exp \left( {{d_1}} \right)$. Based on Lemma 1, using  a similar method as used for $\mathbf{P}_{{1}}$, $\mathbf{P}_{{7}}$ can be solved by iteratively solving $\mathbf{P}_{{8}}$, given as
\begin{subequations}
\begin{align}\label{27}\
&\mathbf{P}_{{8}}: {\mathop {\max }\limits_{{\tau _0},{d _1},{x_k},{f_k},\Upsilon} }\ \Upsilon\\ \notag
&\text{s.t.} \ \ \frac{{T{f_k}}}{C} + \frac{{B\exp \left( {{{{d}}_1}} \right)}}{{{v_k}}}{\log _2}\left( {1 + \frac{{{g_k}\exp \left( {{x_k}} \right)}}{{\sum\limits_{i = k + 1}^K {{g_i}\exp \left( {{x_i}} \right)}  + \sigma _0^2}}} \right)\\
& \ge {R_{k,\min}}, \ 0\leq k\leq K-1,\\
& \frac{{T{f_k}}}{C} + \frac{{B\exp \left( {{d_1}} \right)}}{{{v_k}}}{\log _2}\left( {1 + \frac{{{g_k}\exp \left( {{x_k}} \right)}}{{\sigma _0^2}}} \right)\ge {R_{k,\min}},\\
&{\tau _0}{P_{r,k}} +  \zeta\exp \left( {{{{d}}_1}} \right)\left(\exp \left( {{{{x}}_k}} \right) + {P_{c,k}}\right) + T{\gamma _c}f_k^3 \le {\Phi _k}\left( {{\tau _0},{P_{th}}} \right)\\ \notag
&\frac{{T{f_k}}}{C} + \frac{{B\exp \left( {{{{d}}_1}} \right)}}{{{v_k}}}{\log _2}\left( {1 + \frac{{{g_k}\exp \left( {{x_k}} \right)}}{{\sum\limits_{i = k + 1}^K {{g_i}\exp \left( {{x_i}} \right)}  + \sigma _0^2}}} \right)\\
&- \eta \left[ {{\tau _0}{P_{r,k}} + \zeta\exp \left( {{{{d}}_1}} \right)\left(\exp \left( {{{{x}}_k}} \right) + {P_{c,k}}\right) + T{\gamma _c}f_k^3} \right] \ge   \Upsilon, \\ \notag
&\frac{{T{f_k}}}{C} + \frac{{B\exp \left( {{d_1}} \right)}}{{{v_k}}}{\log _2}\left( {1 + \frac{{{g_k}\exp \left( {{x_k}} \right)}}{{\sigma _0^2}}} \right)\\
&- \eta \left[ {{\tau _0}{P_{r,k}} + \zeta\exp \left( {{{{d}}_1}} \right)\left(\exp \left( {{{{x}}_k}} \right) + {P_{c,k}}\right) + T{\gamma _c}f_k^3} \right] \ge   \Upsilon, \\
& {\tau _0}+\exp \left( {{{{d}}_1}} \right)\leq T,0 \le {\tau _0} \le T,  {d _1} \le \ln\left(T\right), \text{and} \ C6.
\end{align}
\end{subequations}
$\eta$ is a non-negative parameter and $\Upsilon\geq 0$ is an auxiliary variable. In order to address those constraints, auxiliary variables $\exp \left( {{z_k}} \right)$ are introduced. By using SCA, $\mathbf{P}_{{8}}$ can be solved by iteratively solving $\mathbf{P}_{{9}}$.
\begin{subequations}
\begin{align}\label{27}\
&\mathbf{P}_{{9}}: {\mathop {\max }\limits_{{\tau _0},{d _1},{x_k},{z_k}, {f_k},\Upsilon} }\ \Upsilon\\
&\text{s.t.} \ \ \frac{{T{f_k}}}{C} + \exp \left( {{{\overline z }_k^{j}}} \right) + \exp \left( {{{\overline z }_k^{j}}} \right)\left( {{z_k} - {{\overline z }_k^{j}}} \right) \ge {R_{k,\min}},\\
& \frac{{B\exp \left( {{{{d}}_1}} \right)}}{{{v_k}}}{\log _2}\left( {1 + \frac{{{g_k}\exp \left( {{x_k}} \right)}}{{\sum\limits_{i = k + 1}^K {{g_i}\exp \left( {{x_i}} \right)}  + \sigma _0^2}}} \right) \ge \exp \left( {{z_k}} \right),\\
& \frac{{B\exp \left( {{d_1}} \right)}}{{{v_k}}}{\log _2}\left( {1 + \frac{{{g_k}\exp \left( {{x_k}} \right)}}{{\sigma _0^2}}} \right)\ge \exp \left( {{z_k}} \right),k=K,\\
&\frac{{T{f_k}}}{C} + \exp \left( {{{\overline z }_k^{j}}} \right) + \exp \left( {{{\overline z }_k^{j}}} \right)\left( {{z_k} - {{\overline z }_k^{j}}} \right) \\
&- \eta \left[ {{\tau _0}{P_{r,k}} + \zeta\exp \left( {{{{d}}_1}} \right)\left(\exp \left( {{{{x}}_k}} \right) + {P_{c,k}}\right) + T{\gamma _c}f_k^3} \right] \ge   \Upsilon,\\ \notag
& \left(22\rm{d}\right) \ \text{and} \ \left(22\rm{g}\right),
\end{align}
\end{subequations}
where ${\overline z }_k^{j}$, $k\in {\cal K}$, are the given local points at the $j$th iteration. It is not difficult to prove that $\mathbf{P}_{{9}}$ is convex and can be readily solved by using the existing convex optimization tool \cite{S. P. Boyd}.
\begin{table}[htbp]
\begin{center}
\caption{The iterative algorithm based on using SCA}
\begin{tabular}{lcl}
\\\toprule
$\textbf{Algorithm 3}$: The iterative algorithm for $\mathbf{P}_{\text{7}}$\\ \midrule
\  1) \textbf{Input settings:}\\
 \ \ \ \ \ \ \ the error tolerance $\xi_1, \xi_2> 0$, $R_{k,\min} > 0$ and $P_{th}$,\\
  \ \ \ \ \ \ \ the maximum iteration number $N$.\\
\  2) \textbf{Initialization:}\\
 \ \ \ \ \ \ \ EE $\eta _{}^{n} =\eta_0$ and the iteration index $n=0$.\\
\  3) \textbf{Optimization:}\\
\ \ \ \ \ \textbf{$\pmb{\unrhd} $  for \emph{n}=1:N }\\
\ \ \ \ \ \ \ \ \ \ initialize the iterative number $j=1$ and ${\overline z }_k^{j}$;\\
\ \ \ \ \ \ \ \ \ \ \textbf{Repeat:}\\
\ \ \ \ \ \ \ \ \ \ \ \ solve $\text{P}_{\textbf{9}}$ by using \texttt{CVX} for the given $\eta _{}^{n}$;\\
\ \ \ \ \ \ \ \ \ \ \ \  obtain the solution \\
\ \ \ \ \ \ \ \ \ \ \ \  $\left\{{\tau _{0,j}^{*,n}, {d _{1,j}}^{*,n},{x_{k,j}}^{*,n},{z_{k,j}}^{*,n},{f_{k,j}^{*,n}}, \Upsilon_{j}^{*,n}}\right\}$;\\
\ \ \ \ \ \ \ \ \ \ \ \  if $\left| {\Upsilon _j^{ * ,n} - \Upsilon _{j - 1}^{ * ,n}} \right| \le {\xi _2}$\\
\ \ \ \ \ \ \ \ \ \ \ \ \ \  break; \\
\ \ \ \ \ \ \ \ \ \ \ \  else \\
\ \ \ \ \ \ \ \ \ \ \ \ \ \ update $j=j+1$ and ${\overline z }_k^{j}={z_{k,j}}^{*,n}$;\\
\ \ \ \ \ \ \ \ \ \ \ \  end \\
\ \ \ \ \ \ \ \ \ \ \textbf{end Repeat} \\
\ \ \ \ \ \ \ \ \ \  \textbf{if} $\left| \Upsilon_j ^{*,n}-\eta _{}^{n}\right| \le \xi$ \\
\ \ \ \ \ \ \ \ \ \ \ \ \ \ the maximum CE $\Upsilon_j^{*,n}$ is obtained;\\
\ \ \ \ \ \ \ \ \ \ \ \ \ \ break;\\
\ \ \ \ \ \ \ \ \ \  \textbf{else} \\
\ \ \ \ \ \ \ \ \ \ \ \ \ \  update $n=n+1$ and $\eta _{}^{n}=\Upsilon_j ^{*,n}$;\\
\ \ \ \ \ \ \ \ \ \  \textbf{end}\\
\ \ \ \ \ \textbf{$\pmb{\unrhd} $ end} \\
\bottomrule
\end{tabular}
\end{center}
\end{table}

Finally, by iteratively solving $\mathbf{P}_{{9}}$, an iterative algorithm based on SCA is proposed to solve $\mathbf{P}_{{7}}$, which is denoted by Algorithm 3. The details for Algorithm 3 can be found in Table 3. In Algorithm 3, $\xi_1$ and $\xi_2$ are the error tolerances for the CE iteration and the SCA iteration, respectively.
\subsection{ Binary Offloading Mode}
When the binary computation offloading mode is applied, the CE maximization problem  is given as
\begin{subequations}
\begin{align}\label{27}\ \notag
&{\mathop {\max }\limits_{{\tau _0},{\tau _1},{P_k},{P_s},{f_k},\alpha_k} }\ {\mathop {\min }\limits_{k \in {{\cal K}}} }\ \
{\eta _k}\left( {{\tau _0},{\tau _1},{P_k},{f_k},{\alpha _k}} \right) = \\
&\frac{{{\left( {1 - {\alpha_k}} \right)\frac{{T{f_k}}}{C} + {\alpha_k}\frac{{B{\tau _1}}}{{{v_k}}}{{\log }_2}\left( {1 + \frac{{{g_k}{P_k}}}{{\sum\limits_{i = k + 1}^K {{\alpha_i}{g_i}{P_i}}  + \sigma _0^2}}} \right)}}}{{\left( {1 - {\alpha _k}} \right)\left( {{\tau _0}{P_{r,k}} + T{\gamma _c}f_k^3} \right) + {\alpha _k}\left[ {{\tau _0}{P_{r,k}} + {\tau _1}\zeta \left( {{P_k} + {P_{c,k}}} \right)} \right]}}
\\ \notag
&\text{s.t.} \ \ \left( {1 - {\alpha _k}} \right)\frac{{T{f_k}}}{C} + {\alpha _k}\frac{{B{\tau _1}}}{{{v_k}}}{\log _2}\left( {1 + \frac{{{g_k}{P_k}}}{{\sum\limits_{i = k + 1}^K {{\alpha _i}{g_i}{P_i}}  + \sigma _0^2}}} \right)\\
& \geq {R_{k,\min}},k\in {\cal K},\\ \notag
&\left( {1 - {\alpha _k}} \right)\left( {{\tau _0}{P_{r,k}} + T{\gamma _c}f_k^3} \right) + {\alpha _k}\left[ {{\tau _0}{P_{r,k}} + {\tau _1}\zeta \left( {{P_k} + {P_{c,k}}} \right)} \right]\\
&\le {\Phi _k}\left( {{\tau _0},{P_s}} \right),k \in {\cal K},\\
&{\tau _0}{\rm{ + }}\sum\limits_{k = 1}^K {{\alpha _k}{\tau _k}}  \le T, \alpha _k\in\left\{0,\right\}, \left(21\rm{c}\right), \left(21\rm{d}\right), C5, \ \text{and} \ C6,
\end{align}
\end{subequations}
where $\alpha _k$ are the operational mode selection variables for either local computing or complete computation task offloading. $\mathbf{P}_{{10}}$ is a mixed integer non-convex  fractional optimization problem. In order to solve it, motivated by those algorithms for solving $\mathbf{P}_{{4}}$ and $\mathbf{P}_{{7}}$, an alternative algorithm based on SCA can be proposed. Due to the space limit, the details are not presented.  The process iteratively solves $\mathbf{P}_{{11}}$ for the  given $\alpha _k$  and $\eta$ in the following and updates $\alpha _k$ by using Theorem 6.
\begin{subequations}
\begin{align}\label{27}\
&\mathbf{P}_{{11}}: {\mathop {\max }\limits_{{\tau _0},{d _1},{x_k},{z_k}, {f_k},\Upsilon} }\ \Upsilon\\ \notag
&\text{s.t.} \ \ \left( {1 - {\alpha _k}} \right)\frac{{T{f_k}}}{C} + \exp \left( {{{\overline z }_k^{j}}} \right) + \exp \left( {{{\overline z }_k^{j}}} \right)\left( {{z_k} - {{\overline z }_k^{j}}} \right) \\
&\ge {R_{k,\min}},k\in {\cal K},\\
& {\alpha _k}\frac{{B\exp \left( {{{{d}}_1}} \right)}}{{{v_k}}}{\log _2}\left( {1 + \frac{{{g_k}\exp \left( {{x_k}} \right)}}{{\sum\limits_{i = k + 1}^K {{g_i}\exp \left( {{x_i}} \right)}  + \sigma _0^2}}} \right) \ge \exp \left( {{z_k}} \right),\\
& {\alpha _k}\frac{{B\exp \left( {{d_1}} \right)}}{{{v_k}}}{\log _2}\left( {1 + \frac{{{g_k}\exp \left( {{x_k}} \right)}}{{\sigma _0^2}}} \right)\ge \exp \left( {{z_k}} \right),k=K,\\ \notag
&\left( {1 - {\alpha _k}} \right)\frac{{T{f_k}}}{C} + \exp \left( {{{\overline z }_k^{j}}} \right) + \exp \left( {{{\overline z }_k^{j}}} \right)\left( {{z_k} - {{\overline z }_k^{j}}} \right) \\ \notag
& - \eta \left[ {\tau _0}{P_{r,k}} + {\alpha _k}\zeta\exp \left( {{{{d}}_1}} \right)\left(\exp \left( {{{{x}}_k}} \right) + {P_{c,k}}\right) \right.\\
&\left.+ \left( {1 - {\alpha _k}} \right)T{\gamma _c}f_k^3 \right] \ge   \Upsilon, k\in {\cal K},\\ \notag
&{\tau _0}{P_{r,k}} +  {\alpha _k}\zeta\exp \left( {{{{d}}_1}} \right)\left(\exp \left( {{{{x}}_k}} \right) + {P_{c,k}}\right) +\left( {1 - {\alpha _k}} \right) T{\gamma _c}f_k^3 \\
&\le {\Phi _k}\left( {{\tau _0},{P_{th}}} \right),\\
&  \left(22\rm{g}\right)\ \text{and} \ \left(22\rm{h}\right),
\end{align}
\end{subequations}
where ${P_k}{\alpha _k} = \exp \left( {{x_k}} \right)$ and ${\tau _1} = \exp \left( {{d_1}} \right)$. $\Upsilon\geq0$ and $z_k\geq0$ are auxiliary variables. $\eta$ is a non-negative parameter and ${\overline z }_k^{j}$,$k\in {\cal K}$, are the given local points at the $j$th iteration.
\begin{myTheo}
In the wireless powered MEC systems with NOMA under the binary offloading mode, the optimal operational mode selection index has the form given by eq. (26) in the top of the next page,
\begin{figure*}[!t]
\normalsize
\begin{subequations}
\begin{align}\label{27}\
&{\alpha _k}^{*}= \left\{ \begin{array}{l}
0,\text{if}\ {F_{1,k}} < {F_{2,k}},\\
1,\text{otherwise};
\end{array} \right.
\\
&{F_{k,1}} = \left\{ {\begin{array}{*{20}{c}}
\begin{array}{l}
\begin{array}{*{20}{c}}
{\exp \left( {{d_1}} \right)\left[ {\frac{{{\lambda _k}B}}{{{v_k}}}{{\log }_2}\left( {1 + \frac{{{g_k}\exp \left( {{x_k}} \right)}}{{\sigma _0^2}}} \right) - \left( {{\mu _k} + {\omega _k}\eta } \right)\left( {\exp \left( {{x_k}} \right) + {P_{c,k}}} \right)} \right],}&{k = K}
\end{array}\\
\exp \left( {{d_1}} \right)\left[ {\frac{{{\lambda _k}B}}{{{v_k}}}{{\log }_2}\left( {1 + \frac{{{g_k}\exp \left( {{x_k}} \right)}}{{\sum\limits_{i = k + 1}^K {{g_i}\exp \left( {{x_i}} \right)}  + \sigma _0^2}}} \right) - \zeta \left( {{\mu _k} + {\omega _k}\eta } \right)\left( {\exp \left( {{x_k}} \right) + {P_{c,k}}} \right)} \right],\text{otherwise}
\end{array}
\end{array}} \right.\\
&{F_{k,2}} = \left( {{\varpi _{{k}}} + {\omega _k}} \right)\frac{{T{f_k}}}{C} - \left( {{\mu _k} + {\omega _k}\eta } \right)T{\gamma _c}f_k^3,
\end{align}
\end{subequations}
\hrulefill \vspace*{4pt}
\end{figure*}
where $\varpi _{{k}}\geq0$, $\lambda _{{k}}\geq0$, $\omega _k\geq0$ and $\mu _k\geq0$ are the dual variables associated with the constraints given by $\left(25\rm{b}\right)$ - $\left(25\rm{g}\right)$, respectively.
\end{myTheo}

It  can been from Theorem 5 that in the wireless powered MEC networks with NOMA under the binary offloading mode, the optimal operational mode selection also depends on the tradeoff between the achievable computed bits and the energy consumption cost.

Finally, the complexity analysis is presented. Note that there are no references for analyzing the complexity of solving $\mathbf{P}_{\text{9}}$ and $\mathbf{P}_{\text{11}}$ that involves the product of an exponential function and logarithmic function, which involves the division of exponential functions. We cannot provide the complexity analysis for Algorithm 3 and 4. The complexity of Algorithm 1 comes from two parts. One is the for-loop iteration required by using the Dinkelbach's method. Let $L_1$ denote the iteration numbers of the
for-loop. The other is from the solution of $\mathbf{P}_{\text{3}}$ by using \texttt{CVX}. In $\mathbf{P}_{\text{3}}$, there are $3K+2$ variables, $3K+2$ linear matrix inequality (LMI) constraints of size 1, $2K$ third-order inequality constraints and $K$ logarithm  inequality constraints given by $\left(11\rm{b}\right)$. According to the analysis in \cite{S. P. Boyd}-\cite{A. Ben-Tal}, the complexity of Algorithm  1 is ${{\cal O}}\left( n{L_1}\sqrt {9K + 2 + Kn\log \left( n \right)} \left[ \left( {5K + 2} \right) + 2{n^2}K\log \left( n \right) + \right.\right.$ $\left.\left. {n^2} \right] \right)$, where ${\cal O}\left( {\cdot} \right)$ is the big-${\cal O}$ notation and $n={{\cal O}}\left( {3K + 2} \right)$. The complexity of Algorithm 2 are from four parts. Two parts are similar to those of Algorithm 1. The solved problem is $\mathbf{P}_{\text{6}}$ instead of $\mathbf{P}_{\text{3}}$. The third part is from the subgradient method and the fourth part is from the alternative optimization. Let $L_2$ and $L_3$ denote the iteration numbers of the for-loop part and that of alternative optimization, respectively. Let $\ell_1$ denote the tolerance error for the subgradient method. Similar to the analysis for Algorithm 1, the complexity of Algorithm 2 is  ${{\cal O}}\left( n{L_2}{L_3}\left[\sqrt {9K + 2 + Kn\log \left( n \right)} \left[ \left( {5K + 2} \right) + 2{n^2}K\log \right.\right.\right.\\ \left.\left.\left.\left( n \right) + {n^2} \right] +1/\ell_2^2\right]\right)$.
\section{Simulation Results}
\begin{table}[htbp]
\centering
 \caption{\label{tab:test}Simulation Parameters}
 \begin{tabular}{l|c|c}
  \midrule
  \midrule
  Parameters & Notation & Typical Values  \\
  \midrule
  \midrule
 Numbers of Users & $K$ & $5$ \\
The maximum EH power & $P_{k}^{\max}$ & $0.004927$ W\\
 The sensitivity threshold & $P_0$ & $0.000064 $ W\\
The  communication bandwidth & $B$ & $2 $  MHz\\
 Circuit parameter & $\mu$ & $274$ \\
 Circuit parameter & $\psi$ & $0.29$ \\
  The noise power & $\sigma_0^{2}$ & $10^{-9}$ W \\
 The number of cycles for one bit& $C$ & $10^3$ cycles/bit \\
 The  capacitance coefficient & $\gamma _{c}$ & $10^{-28}$ \\
 The tolerance error & $\varpi$ & $10^{-4}$ \\
  The minimum computation bits& $R_{k,\min}$ & $10^{4}$ Bits\\
   The amplifier coefficient& $\zeta$ & $3$\\
   The received power& $P_{r,k}$ & $5$ dbm\\
   The constant circuit power& $P_{c,k}$ & $5$ dbm\\
\midrule
\midrule
 \end{tabular}
\end{table}
In this section, simulation results are presented to evaluate the  proposed CE maximization framework and compare its performance with the existing computation bits  (CB) maximization framework. The simulation parameters are selected  based on the works in \cite{F. Wang}, \cite{S. Bi} and the parameters for the non-linear EH model are selected based on \cite{S. Wang}. Similar to \cite{Q. Wu3}-\cite{K. Chi}, the reference distance is set as 1 meter and the maximum services distance for users is 5 meters. The channel power gains are set the same as those in \cite{Q. Wu3}.  The details for the parameters are given in Table IV.

\begin{figure}[htb]
\centering
\includegraphics[width=3.2 in,height=2.5 in]{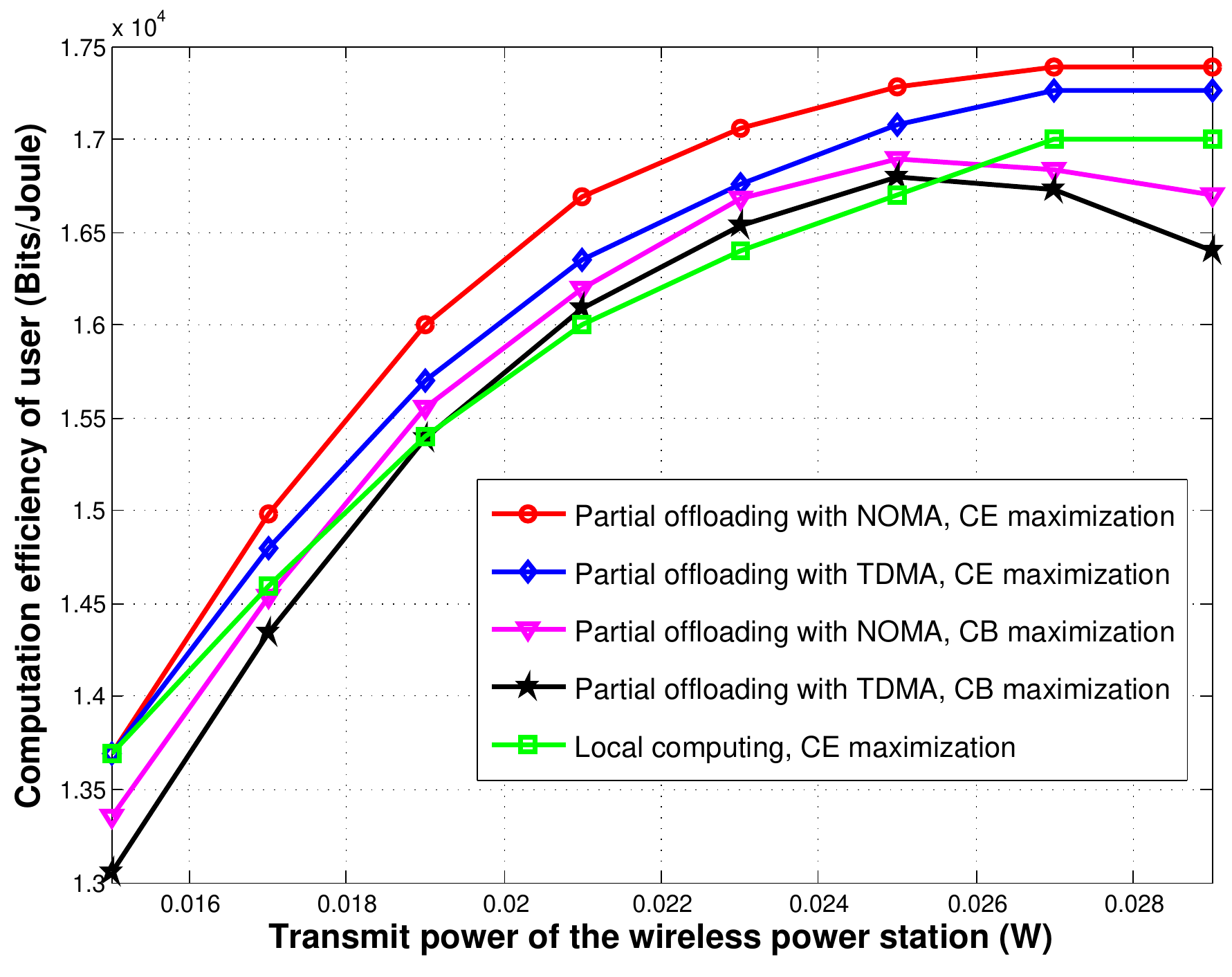}
\includegraphics[width=3.2 in,height=2.5 in]{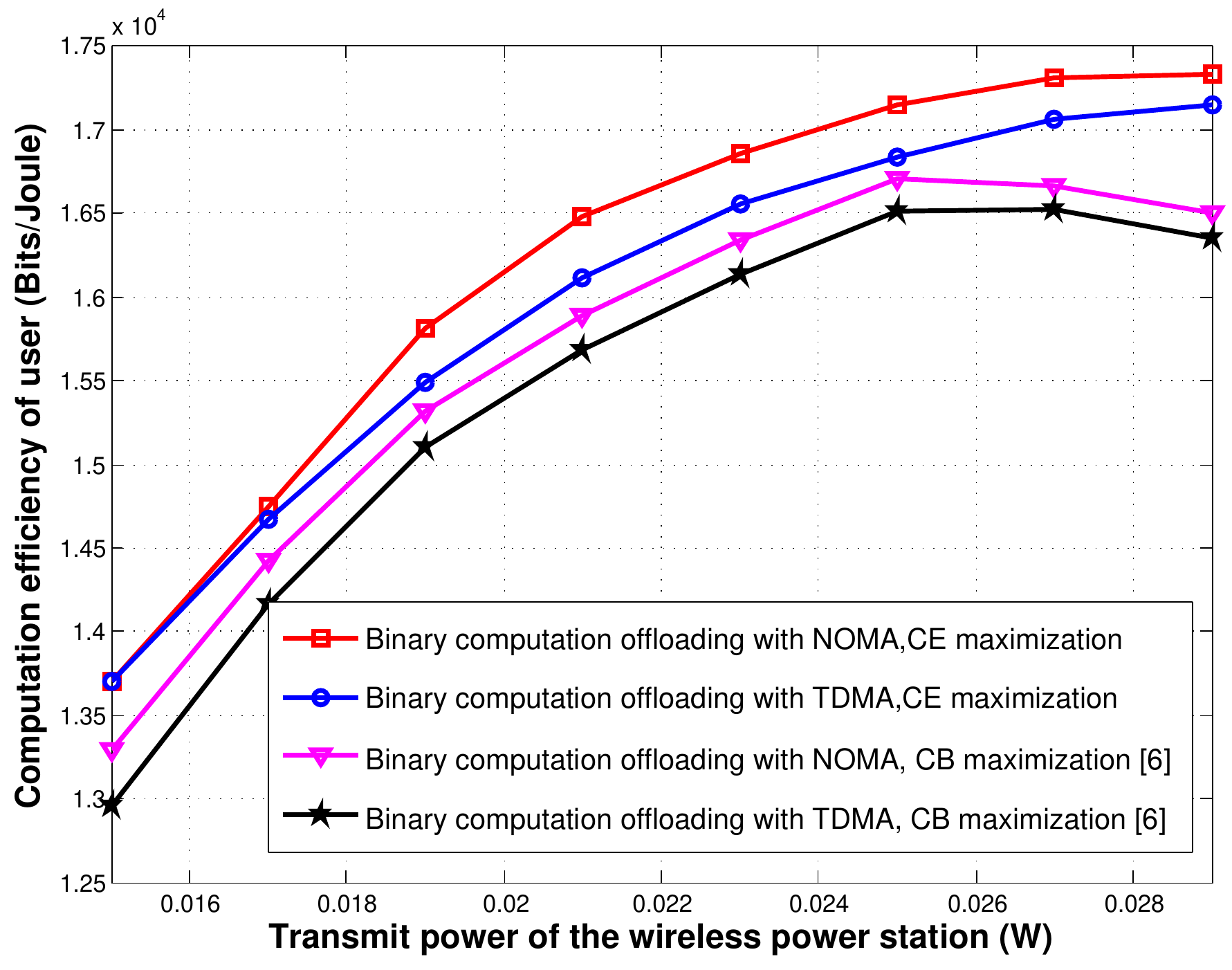}
\put(-308,-10){\footnotesize{(a)}}
\put(-100,-10){\footnotesize{(b)}}
\caption{(a) CE  under partial offloading versus the transmission power of the wireless power station; (b) CE under binary offloading versus the transmission power of the wireless power station.} \label{fig1}
\end{figure}
Fig. 3 shows the CE achieved with the max-min fairness criterion versus the transmission power of the wireless power station using the proposed CE maximization framework and the CB maximization framework  under both partial and binary  offloading modes. The CB maximization framework is to maximize the number of computation bits under the max-min fairness criterion. It is seen that the CB maximization framework cannot guarantee that the maximum CE can be achieved. This indicates that the resource allocation schemes for maximizing the number of CB are inappropriate to the wireless powered MEC network that aim to achieve the maximum CE. Moreover, on one hand,  in the CB maximization framework,  the CE firstly increases with the transmission power and then decreases when the transmit power is large enough. On the other hand,  the number of CB always  increases with the transmission power. Thus, it is found that there is a tradeoff between the CE and the number of CB. It  can also be seen that the CE achieved with NOMA is larger than that obtained with TDMA, irrespective of the selected offloading mode. This indicates that NOMA can obtain a CE gain compared to TDMA. The reason is that the offloading efficiency is higher when NOMA is applied compared to that of TDMA  \cite{Z. Ding}, \cite{A. Kiani}.

\begin{figure}[!t]
\centering
\includegraphics[width=3.5 in]{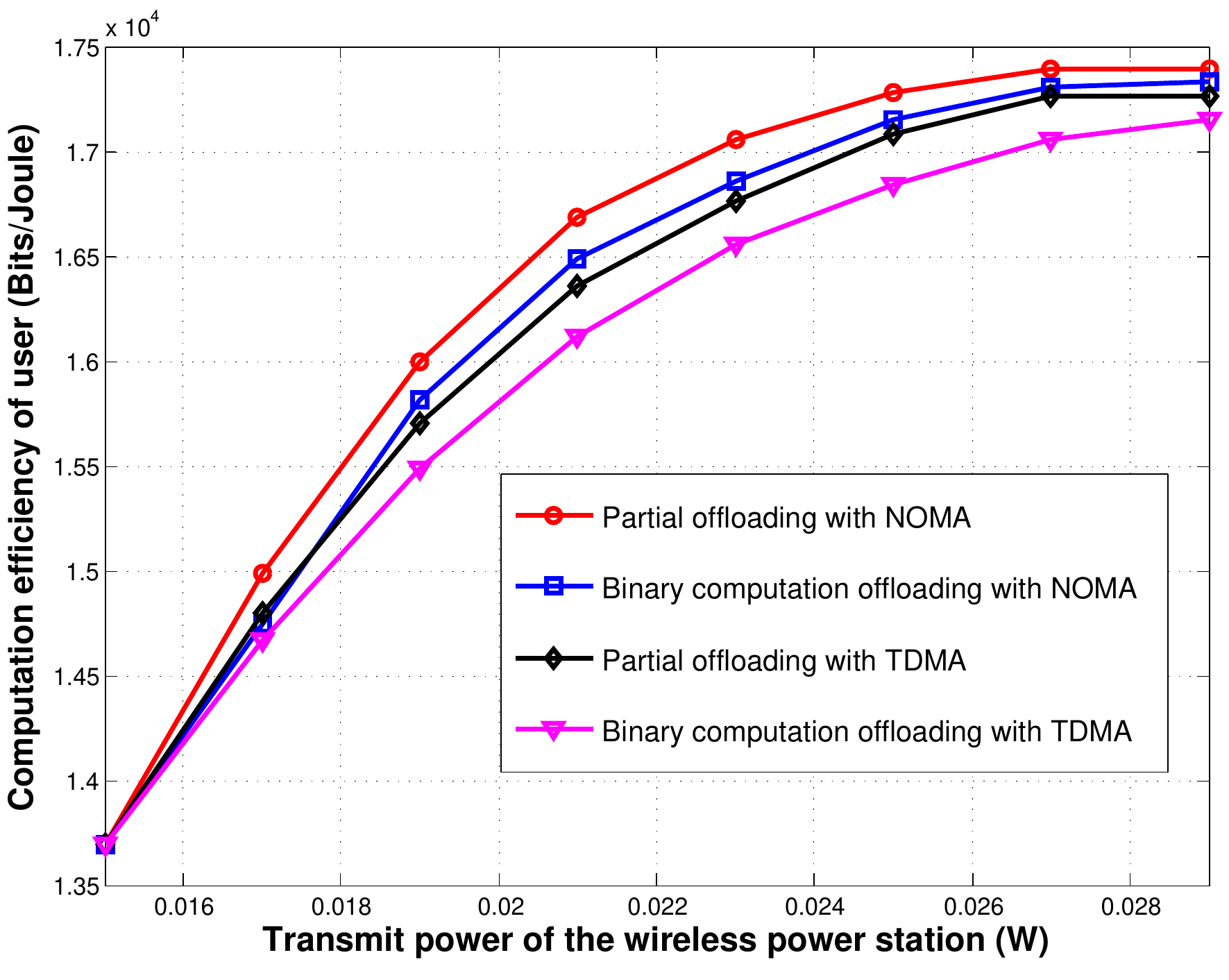}
\caption{The CE comparison achieved under different offloading modes and multiple access schemes.} \label{fig.1}
\end{figure}
Fig. 4 comprehensively presents the CE comparison achieved under different operation modes and different multiple access schemes. It can be seen that CE achieved under all the cases is the same when the transmit power of the wireless power station is small. The reason is that when the transmit power of the wireless power station is small, all the users choose to perform local computing even under the partial computation offloading mode due to the fact that  the harvested energy is very small. This is consistent with our theoretical analysis presented in Section III. It can be also seen that the CE achieved under the partial offloading mode is larger than that obtained under the binary offloading mode, irrespective of the multiple access schemes. The reason is that the partial offloading mode can flexibly allocate resources for computation offloading and local computing while the resources under the binary offloading mode can only be completely allocated either for local computing or for computation offloading.

\begin{figure}[htb]
\centering
\includegraphics[width=3.2 in,height=2.5 in]{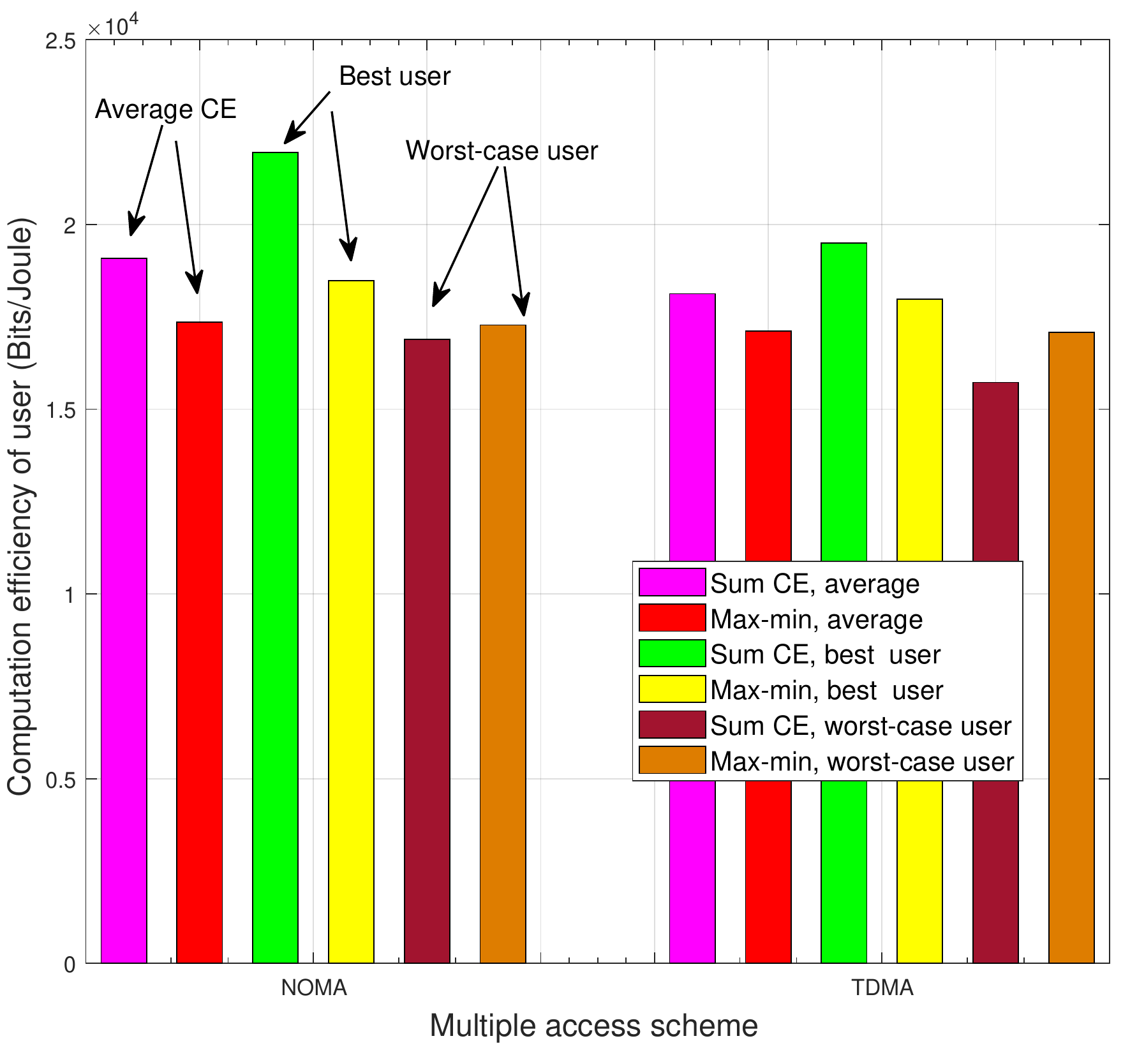}
\includegraphics[width=3.2 in,height=2.5 in]{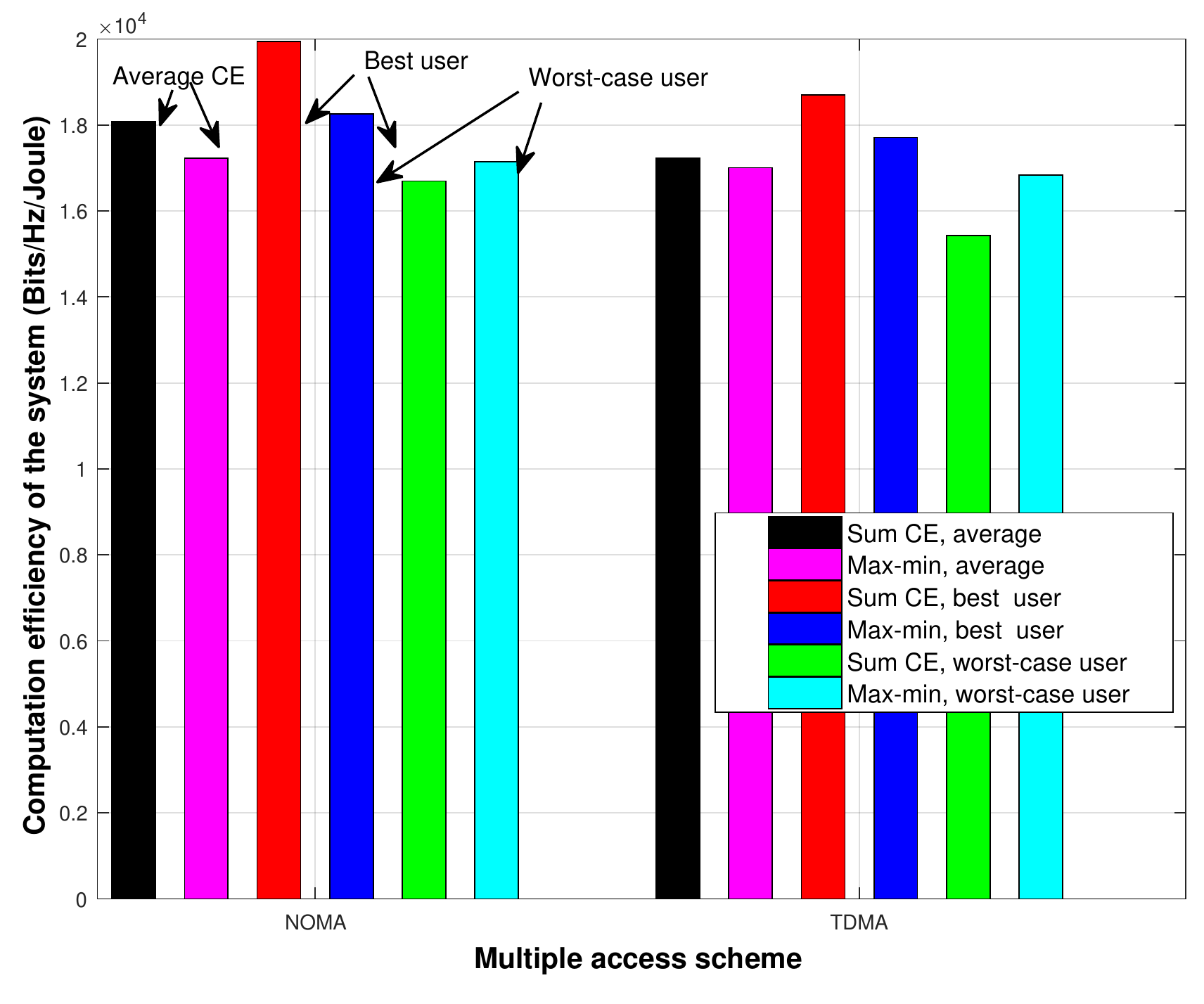}
\put(-308,-10){\footnotesize{(a)}}
\put(-100,-10){\footnotesize{(b)}}
\caption{(a) Fairness comparison under partial offloading mode with different optimization objectives; (b) Fairness comparison under binary offloading mode with different optimization objectives.} \label{fig1}
\end{figure}
Fig. 5 compares the user fairness achieved with our proposed max-min fairness criterion framework and the sum CE maximization framework under both the partial and binary computation offloading modes. Note that the sum CE maximization framework is to maximize the sum of CE of all users. The transmission power of the wireless power station is $0.025$ W. It can be seen that there is a tradeoff between the sum CE and the fairness among users. The max-min fairness criterion framework can improve the fairness among users at the cost of the sum CE. The reason is that our proposed resource allocation schemes aim to maximize the minimum CE among all users while those schemes for maximizing the sum CE allocate more resource to the user with a better offloading efficiency.

\begin{figure}[!t]
\centering
\includegraphics[width=3.5 in]{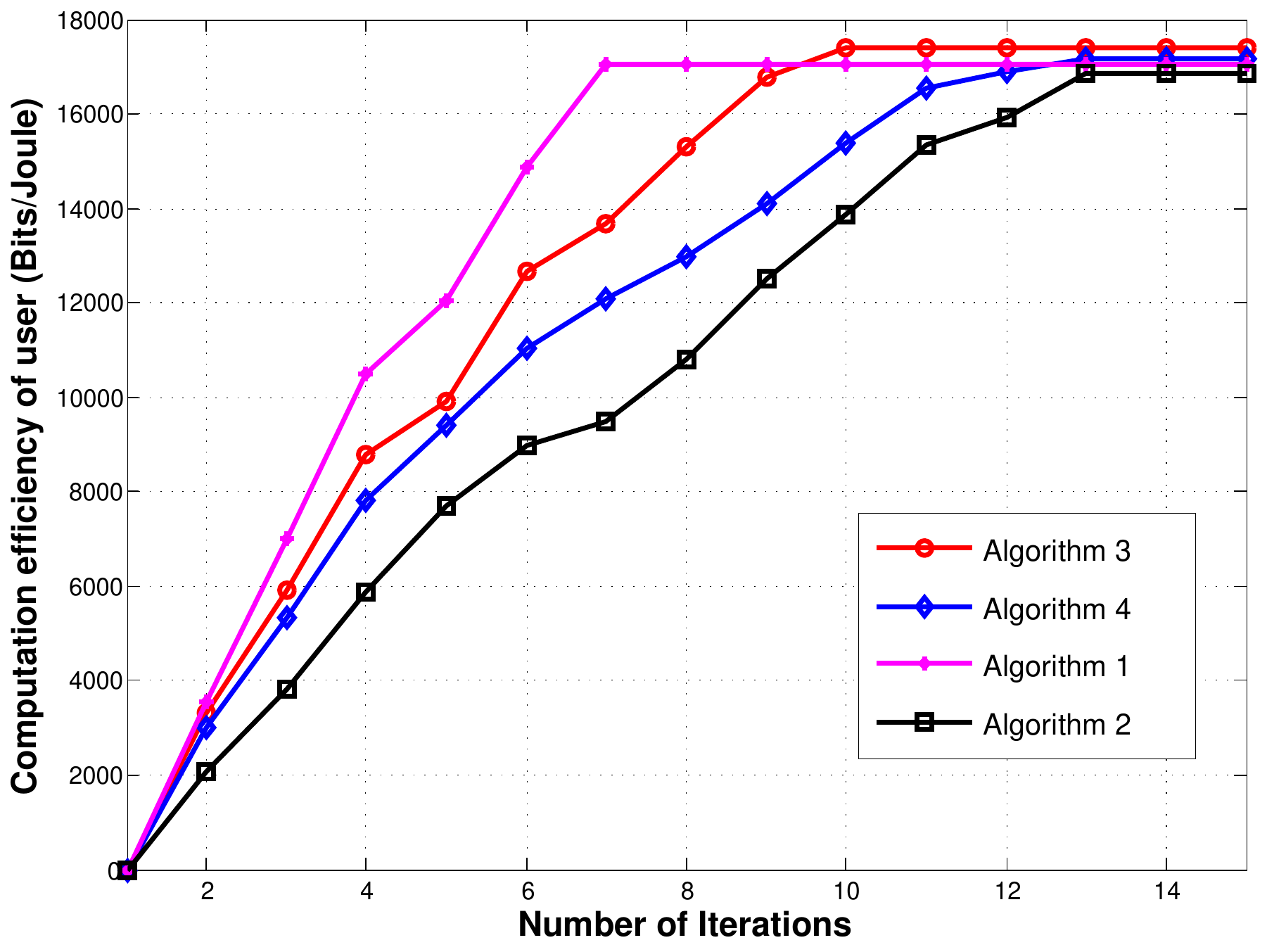}
\caption{CE value versus the number of iterations} \label{fig.1}
\end{figure}
Fig. 6 shows the CE versus the number of iterations of different algorithms. It can be  seen that less than 15 iterations are required for all algorithms to converge to the maximum CE. This indicates that our proposed algorithm is computationally efficient. Moreover, it can be seen that the number of iterations of Algorithm 1 is less than that of other algorithms.  It only needs to update the CE  while other algorithms need to update the operational mode selection variable or perform SCA iteration.
\section{Conclusions}
In this paper a new performance metric called CE was defined and  studied  in a wireless powered MEC network under both  partial and binary offloading modes.  TDMA and NOMA were investigated for offloading transmission under a practical non-linear EH model. The EH time, the CPU frequencies, the user offloading times, and the user transmit powers were jointly optimized to maximize the CE under the max-min  fairness criterion. Two iterative algorithms and two alternative optimization algorithms were proposed to tackle those challenging non-convex optimization problems. It was shown that our proposed resource allocation strategies are superior to other benchmark schemes in terms of CE.  It was also shown that the CE achieved with the partial computation offloading mode outperforms that obtained with the binary computation offloading mode and NOMA can always achieve a  CE gain compared to TDMA. The study also elucidated the performance  tradeoff between CE and the CB.
\appendices
\section{Proof of Theorem 2}
Let $\lambda _k\geq0$, $\rho _k\geq0$, $\theta _k\geq0$, and $\beta\geq0$ denote the dual variables corresponding to the constraints given by $\left(11\rm{b}\right)$-$\left(11\rm{d}\right)$ and the constraint C3,  respectively. Then, the Lagrangian of $\mathbf{P}_{{3}}$  can be given as
\begin{align}\label{27}\ \notag
&{\cal{L}}\left( \Xi  \right) =  \sum\limits_{k = 1}^K {{\lambda _k}} \left( {\frac{{T{f_k}}}{C} + \frac{{B{\tau _k}}}{{{v_k}}}{{\log }_2}\left( {1 + \frac{{{g_k}{y_k}}}{{{\tau _k}\sigma _0^2}}} \right) - {R_{k,{\rm{min}}}}} \right)\\ \notag
 &+ \sum\limits_{k = 1}^K {{\rho _k}} \left( {{\Phi _k}\left( {{\tau _0},{P_{th}}} \right) - {\tau _0}{P_{r,k}} - \zeta{y_k} - \zeta{\tau _k}{P_{c,k}} - T{\gamma _c}f_k^3} \right)\\ \notag
 &+ \beta \left( {T - \sum\limits_{n = 0}^K {{\tau _n}} } \right)+ \sum\limits_{k = 1}^K {{\theta _k}} \left( \frac{{T{f_k}}}{C} + \frac{{B{\tau _k}}}{{{v_k}}}{{\log }_2}\left( {1 + \frac{{{g_k}{y_k}}}{{{\tau _k}\sigma _0^2}}} \right)\right.\\
  &\left.- \eta \left[ {{\tau _0}{P_{r,k}} + \zeta{y_k} + \zeta{\tau _k}{P_{c,k}} + T{\gamma _c}f_k^3} \right] - \Upsilon  \right)+\Upsilon,
\end{align}
where $\Xi $ denotes a collection of all the primal and dual variables related to $\mathbf{P}_{{3}}$.

Based on the Lagrangian of $\mathbf{P}_{{3}}$, the derivations of the Lagrangian with respect to $f_k$ and $y_k$, can be respectively given as
\begin{subequations}
\begin{align}\label{27}\
&\frac{{\partial {\cal{L}}\left( \Xi  \right)}}{{{f_k}}} = \frac{{T{\lambda _k}}}{C} - 3{\rho _k}T{\gamma _c}f_k^2 + {\theta _k}\left( {\frac{T}{C} - \eta 3T{\gamma _c}f_k^2} \right),\\
&\frac{{\partial {\cal{L}}\left( \Xi  \right)}}{{{y_k}}} = \frac{{\left( {{\theta _k} + {\lambda _k}} \right)B{\tau _k}{g_k}}}{{{v_k}\ln 2\left( {{\tau _k}\sigma _0^2 + {g_k}{y_k}} \right)}} - \zeta\left( {{\theta _k}\eta  + {\rho _k}} \right).
\end{align}
\end{subequations}
Let their derivations be zero. Thus, $\left(12\rm{a}\right)$ can be obtained, and one has
\begin{align}\label{27}\
{y_k} = {\left[ {\frac{{\left( {{\lambda _k} + {\theta _k}} \right)B{\tau _k}}}{{\zeta{v_k}\ln 2\left( {{\rho _k} + {\theta _k}\eta } \right)}} - \frac{{{\tau _k}\sigma _0^2}}{{{g_k}}}} \right]^ + }.
\end{align}
When $\tau _k=0$, it is not difficult  to prove that $P_k^{opt}=0$. Since $y_k=\tau{ _k}P_k, k\in {\cal K}$, when $\tau _k\neq0$, $P_k^{opt}=y_k/\tau{ _k}$. Thus, $\left(12\rm{b}\right)$  is obtained. The proof for Theorem 2 is complete.
\section{Proof of Theorem 3}
Let $z\left( {{\rho _k},\beta ,{\theta _k}} \right)$ and $\Gamma\left( {\lambda _k, {\rho _k},\beta ,{\theta _k}, g _k} \right)$ denote the derivation of the Lagrangian ${\cal{L}}\left( \Xi  \right)$ with respect to $\tau_0$ and $\tau_k$, respectively. They are respectively  expressed as
\begin{subequations}
\begin{align}\label{27}\
&z\left( {{\rho _k},\beta ,{\theta _k}} \right)= \sum\limits_{k = 1}^K {{\rho _k}} \left( {{P_{E,k}} - {P_{r,k}}} \right) - \beta  - \sum\limits_{k = 1}^K {{\theta _k}\eta } {P_{r,k}},\\ \notag
&\Gamma\left( {\lambda _k, {\rho _k},\beta ,{\theta _k},g _k} \right) = \\ \notag
&\frac{\left( {{\lambda _k} + {\theta _k}} \right)B}{{{v_k}}}\left[ {{{\log }_2}\left( {1 + \frac{{{g_k}{P_k}}}{{\sigma _0^2}}} \right) - \frac{{{g_k}{P_k}}}{{\ln 2\left( {\sigma _0^2 + {g_k}{P_k}} \right)}}} \right]  \\
&- \zeta\left( {{\rho _k} + {\theta _k}\eta } \right){P_{c,k}} - \beta.
\end{align}
\end{subequations}
For the given $\lambda _k$, $\rho _k$, $\theta _k$ and $\beta$, it can be seen from $\left(26\right)$ that ${\cal{L}}\left( \Xi  \right)$ is a linear function of $\tau_0$ and $\tau_k$. Since $\mathbf{P}_{{3}}$ is convex and the Slater's conditions are satisfied, ${\cal{L}}\left( \Xi  \right)$ is upper-bounded with respect to $\tau_0$ and $\tau_k$. Thus, $z\left( {{\rho _k},\beta ,{\theta _k}} \right)\leq0$. When $z\left( {{\rho _k},\beta ,{\theta _k}} \right)<0$, the maximum of ${\cal{L}}\left( \Xi  \right)$ is achieved when $\tau_0=0$. When $z\left( {{\rho _k},\beta ,{\theta _k}} \right)=0$, $\tau_0$ can be any  arbitrary value within $\left[ {0,T} \right)$ since ${\cal{L}}\left( \Xi  \right)$ is constant with respect to $\tau_0$. Thus, $\left(13\rm{a}\right)$  is obtained.

By using $\left(29\rm{b}\right)$ and substituting $\left(12\rm{b}\right)$ into $\left(30\rm{b}\right)$, one has
\begin{align}\label{27}\ \notag
&\Gamma\left( {\lambda _k, {\rho _k},\beta ,{\theta _k},g _k} \right)=\\ \notag
&\frac{ \left( {{\lambda _k} + {\theta _k}} \right)B}{{{v_k}}}\left[ {{{\log }_2}\left( {1 + \frac{{{g_k}}}{{\sigma _0^2}}{{\left[ {\frac{{\left( {{\lambda _k} + {\theta _k}} \right)B}}{{\zeta{v_k}\ln 2\left( {{\rho _k} + {\theta _k}\eta } \right)}} - \frac{{\sigma _0^2}}{{{g_k}}}} \right]}^ + }} \right)} \right] \\
&- \zeta\left( {{\rho _k} + {\theta _k}\eta } \right)\left( {{{\left[ {\frac{{\left( {{\lambda _k} + {\theta _k}} \right)B}}{{\zeta{v_k}\ln 2\left( {{\rho _k} + {\theta _k}\eta } \right)}} - \frac{{\sigma _0^2}}{{{g_k}}}} \right]}^ + } + {P_{c,k}}} \right) - \beta.
\end{align}
It can be seen from $\left(30\right)$ that $\Gamma\left( {\lambda _k, {\rho _k},\beta ,{\theta _k}, g _k} \right)$ increases with $g_k$ and decreases with $\rho _k$ when $g_k\geq {\left[ {\sigma _0^2\zeta{v_k}\ln 2\left( {{\rho _k} + {\theta _k}\eta } \right)} \right]}/{\left( {{\lambda _k} + {\theta _k}} \right)B}$. Similar to the derivation for $\left(13\rm{a}\right)$, since $\tau_k\geq0$, one has $\tau_k=0$ when $\Gamma\left( {\lambda _k, {\rho _k},\beta ,{\theta _k},g _k} \right)<0$ and $\tau_k\geq0$ when $\Gamma\left( {\lambda _k, {\rho _k},\beta ,{\theta _k},g _k} \right)=0$. Since $\rho _k\geq0$, $\Gamma\left( {\lambda _k, {\rho _k},\beta ,{\theta _k}, g _k} \right)$ achieves its maximum when $\rho _k=0$. Moreover, $\Gamma\left( {\lambda _k, 0,\beta ,{\theta _k}, g _k} \right)= - \zeta\left( {{\rho _k} + {\theta _k}\eta } \right) {P_{c,k}} - \beta<0$ when ${g_k} = \frac{{\sigma _0^2\zeta{v_k}{\theta _k}\eta \ln 2}}{{\left( {{\lambda _k} + {\theta _k}} \right)B}}$, and $\Gamma\left( {\lambda _k, 0,\beta ,{\theta _k}, g _k} \right)$ tends to $+\infty$ when  $g_k$ goes to $+\infty$. This indicates that $\Gamma\left( {\lambda _k, {0},\beta ,{\theta _k}, g _k} \right)$ always has $w^{opt}$ that satisfies $\Gamma\left( {\lambda _k, {0},\beta ,{\theta _k}, w^{opt}} \right)=0$. Thus, $\left(15\right)$ is proved. Based on $\left(30\right)$ and $\left(15\right)$, the following cases can be obtained. When $g_k<w^{opt}$,   $t_k=0$ since  $\Gamma\left( {\lambda _k, {\rho _k},\beta ,{\theta _k}, g_k} \right)<0$; when $g_k>w^{opt}$, $\Gamma\left( {\lambda _k, {\rho _k},\beta ,{\theta _k}, g_k} \right)<0$ holds when ${\rho _k}>0$. In this case, $t_k=0$ and ${\tau _0}{P_{r,k}} + \zeta{\tau _k}\left({P_k} +{P_{c,k}}\right)+ T{\gamma _c}f_k^3 < {\Phi _k}\left( {{\tau _0},{P_s}} \right)$. This contradicts the complementary slackness condition that ${{\rho _k}} \left( {{\Phi _k}\left( {{\tau _0},{P_{th}}} \right) - {\tau _0}{P_{r,k}} - \zeta{\tau _k}\left({P_k} +{P_{c,k}}\right) - T{\gamma _c}f_k^3} \right)=0$. Thus, when $g_k>w^{opt}$ and $\rho _k>0$, one has $ {{\Phi _k}\left( {{\tau _0},{P_{th}}} \right) - {\tau _0}{P_{r,k}} - \zeta{\tau _k}\left({P_k} +{P_{c,k}}\right) - T{\gamma _c}f_k^3} =0$. For $g_k=w^{opt}$, since $\Gamma\left( {\lambda _k, {\rho _k},\beta ,{\theta _k}, g_k} \right)<\Gamma\left( {\lambda _k, {0},\beta ,{\theta _k}, g_k} \right)=0$ when $\rho _k>0$, $\tau _k=0$. In this case, the complementary slackness condition that ${{\rho _k}} \left( {{\Phi _k}\left( {{\tau _0},{P_{th}}} \right) - {\tau _0}{P_{r,k}} - \zeta{\tau _k}\left({P_k} +{P_{c,k}}\right) - T{\gamma _c}f_k^3} \right)=0$ cannot be held. Thus, one has $\rho _k=0$ and ${{\rho _k}} \left( {{\Phi _k}\left( {{\tau _0},{P_{th}}} \right) - {\tau _0}{P_{r,k}} - \zeta{\tau _k}\left({P_k} +{P_{c,k}}\right) - T{\gamma _c}f_k^3} \right)\leq0$. From the above analysis, $\left(14\right)$ is obtained. Thus, the proof for Theorem 3 is complete.

\begin{IEEEbiography}[{\includegraphics[width=1.0in,height=1.15in,clip,keepaspectratio]{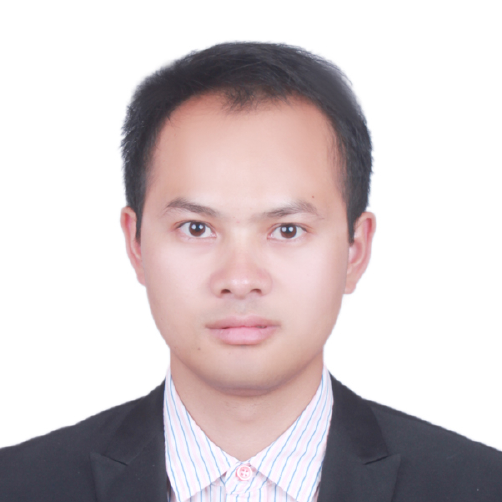}}]{Fuhui Zhou}

has worked as a Senior Research Fellow at Utah State University. He received the Ph. D. degree from Xidian University, Xian, China, in 2016. He is currently a Full Professor at College of Electronic and Information Engineering, Nanjing University of Aeronautics and Astronautics. His research interests focus on cognitive radio, edge computing, machine learning, NOMA, physical layer security, and resource allocation. He has published more than 90 papers, including IEEE Journal of Selected Areas in Communications, IEEE Transactions on Wireless Communications, IEEE Wireless Communications, IEEE Network, IEEE GLOBECOM, etc. He was awarded as Young Elite Scientist Award of China. He has served as Technical Program Committee (TPC) member for many International conferences, such as IEEE GLOBECOM, IEEE ICC, etc. He serves as an Associate Editor of IEEE Transactions on Communications, IEEE Systems Journal, IEEE Wireless Communications Letters, IEEE Access and Physical Communications. He also serves as co-chair of IEEE Globecom 2019 and IEEE ICC 2019 workshop on Advanced Mobile Edge /Fog Computing for 5G Mobile Networks and Beyond.
\end{IEEEbiography}

\begin{IEEEbiography}[{\includegraphics[width=1.0in,height=1.15in,clip,keepaspectratio]{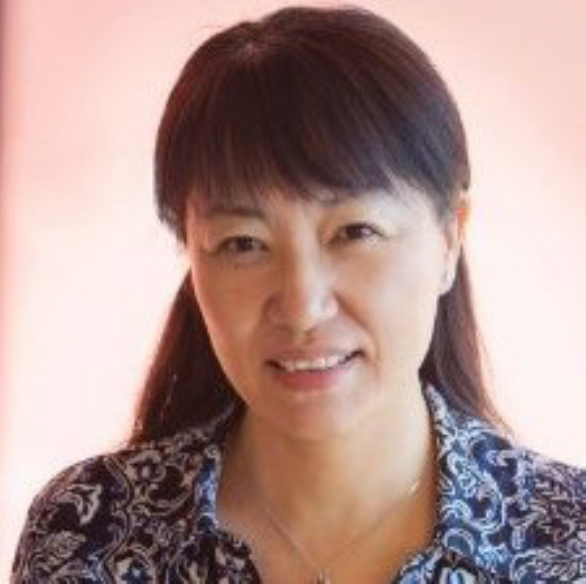}}]{Rose Qingyang Hu}

 (S'95-M'98-SM'06-F'20)  is a Professor in the Electrical and Computer Engineering Department and Associate Dean for research of College of Engineering  at Utah State University. She also directs Communications Network Innovation Lab at Utah State University. Her current research interests include next-generation wireless system design and optimization, Internet of Things, Cyber Physical system, Mobile Edge Computing, V2X communications, artificial intelligence in wireless networks, wireless system modeling and performance analysis. Prof. Hu received the B.S. degree from the University of Science and Technology of China, the M.S. degree from New York University, and the Ph.D. degree from the University of Kansas.  Besides a decade academia experience, she has more than 10 years of R\&D experience with Nortel, Blackberry, and Intel as a Technical Manager, a Senior Wireless System Architect, and a Senior Research Scientist, actively participating in industrial 3G/4G technology development, standardization, system level simulation, and performance evaluation.   She has published extensively and holds numerous patents in her research areas.  Prof. Hu is currently serving on the editorial boards of  the IEEE Transactions on Wireless Communications, the IEEE Transactions on Vehicular Technology, the IEEE Communications Magazine and the IEEE Wireless Communications.  She also served as the TPC Co-Chair for the IEEE ICC 2018. She is an IEEE Communications Society Distinguished Lecturer Class 2015-2018.  She was a recipient of prestigious Best Paper Awards from the IEEE GLOBECOM 2012, the IEEE ICC 2015, the IEEE VTC Spring 2016, and the IEEE ICC 2016.  Prof. Hu is fellow of IEEE and a member of Phi Kappa Phi Honor Society.

\end{IEEEbiography}
\end{document}